\documentclass[
	journal
]{IEEEtran}

\IEEEoverridecommandlockouts 
\makeatletter
\def\markboth#1#2{\def\leftmark{\@IEEEcompsoconly{\sffamily}\MakeUppercase{\protect#1}}%
\def\rightmark{\@IEEEcompsoconly{\sffamily}\MakeUppercase{\protect#2}}}
\makeatother

\usepackage[amssymb]{SIunits}
\usepackage[cmex10]{amsmath}
\usepackage{amssymb}
\usepackage{amsthm}
\usepackage[latin1]{inputenc}
\usepackage{cite}
\usepackage[pdftex]{graphicx}
\usepackage{url}
\usepackage{epstopdf}
\usepackage{glossaries}
\usepackage{graphicx}
\usepackage[small,bf]{caption}
\usepackage[subrefformat=parens,labelformat=parens]{subfig}
\usepackage{bm}
\usepackage[usenames]{color}
\usepackage[lined,ruled,linesnumbered]{algorithm2e}
\usepackage{balance}
\usepackage{array}
\usepackage{bbm}
\usepackage{ctable}

\usepackage{mathtools}

\newcommand\demo{\xqed{$\triangle$}}

\newcommand\coolunder[2]{\mathrlap{\smash{\underbrace{\phantom{%
    \begin{matrix} #2 \end{matrix}}}_{\mbox{$#1$}}}}#2}
    
\newcommand{\otoprule}{\midrule[\heavyrulewidth]}

\newcommand{\xqed}[1]{%
  \leavevmode\unskip\penalty9999 \hbox{}\nobreak\hfill
  \quad\hbox{#1}}

\newtheorem{example}{Example}

\newcolumntype{L}[1]{>{\raggedright\let\newline\\\arraybackslash\hspace{0pt}}m{#1}}
\newcolumntype{C}[1]{>{\centering\let\newline\\\arraybackslash\hspace{0pt}}m{#1}}
\newcolumntype{R}[1]{>{\raggedleft\let\newline\\\arraybackslash\hspace{0pt}}m{#1}}

\graphicspath{%
	{../pstricks/}%
}%

\newcommand{\GaussBin}{G}

\def\thBP{\varepsilon^{\mathrm{BP}}}
\def\thMAP{\varepsilon^{\mathrm{MAP}}}
\def\dl{d_{\mathrm v}}
\def\dr{d_{\mathrm c}}
\def\pvec{{\boldsymbol p}}
\def\xvec{{\boldsymbol x}}
\def\evec{{\boldsymbol e}}
\def\avec{{\boldsymbol a}}
\def\bvec{{\boldsymbol b}}
\def\cvec{{\boldsymbol c}}
\def\avecc{{\boldsymbol a}_\circ}
\def\bvecc{{\boldsymbol b}_\circ}

\def\Fvec{{\boldsymbol F}}
\def\Gvec{{\boldsymbol G}}
\def\Xvec{{\boldsymbol X}}
\def\Yvec{{\boldsymbol Y}}
\def\yvec{{\boldsymbol y}}
\def\zvec{{\boldsymbol z}}
\def\Dvec{{\boldsymbol D}}
\def\Avec{{\boldsymbol A}}
\def\Bvec{{\boldsymbol B}}

\def\fvec{{\boldsymbol f}}
\def\hvec{{\boldsymbol h}}
\def\gvec{{\boldsymbol g}}
\def\H{{\boldsymbol H}}

\def\fvecy{{\boldsymbol f}(\yvec;\varepsilon)}
\def\gvecx{{\boldsymbol g}(\xvec)}

\def\vvec{{\boldsymbol v}}
\def\svec{{\boldsymbol s}}
\def\zerovec{{\boldsymbol 0}}
\def\onevec{{\boldsymbol 1}}
\def\pvecO{{\boldsymbol p}_{\circ}}
\def\xvecO{{\boldsymbol x}_{\circ}}
\def\yvecO{{\boldsymbol y}_{\circ}}
\def\fvecO{{\boldsymbol f}_{\circ}}
\def\gvecO{{\boldsymbol g}_{\circ}}

\newcommand{\W}{\mathbf{W}}
\newcommand{\U}{\mathbf{U}}
\newcommand{\V}{\mathbf{V}}

\newcommand{\eps}{\varepsilon}

\newcommand{\set}[2]{\mathcal{S}^{#1}_{#2}}
\newcommand{\setg}[1]{\mathcal{S}^{g}_{#1}}
\newcommand{\setf}[1]{\mathcal{S}^{f}_{#1}}

\newcommand{\deriv}[1]{\frac{\partial}{\partial #1}}
\newcommand{\derivF}[2]{\frac{\partial #1}{\partial #2}}

\def\Egap{{\Delta E}}
\newcommand{\T}{^{\mathsf{T}}}

\def\fvecy{{\boldsymbol f}(\yvec;\varepsilon)}
\def\gvecx{{\boldsymbol g}(\xvec)}
\def\Fvecy{{F}(\yvec;\varepsilon)}
\def\Gvecx{{G}(\xvec)}
\def\dw{{w}}

\def\tr{{\mathrm{Tr}}}

\newtheorem{lemma}{Lemma}
\newtheorem{theorem}{Theorem}
\newtheorem{proposition}{Proposition}
\newtheorem{definition}{Definition}
\newtheorem{corollary}{Corollary}
\newtheorem{remark}{Remark}

\begin{document}

\title{Threshold Saturation for Nonbinary SC-LDPC Codes on the Binary Erasure Channel}

\author{
	\IEEEauthorblockN{
	Iryna Andriyanova, \emph{Member, IEEE}, and
	Alexandre Graell i Amat, \emph{Senior Member, IEEE}
	}
}



\maketitle
\begin{abstract}
We analyze the asymptotic performance of nonbinary spatially-coupled low-density parity-check (SC-LDPC) code ensembles defined over the general linear group on the binary erasure channel. In particular, we prove threshold saturation of belief propagation decoding to the 
so called potential threshold, using the proof technique based on potential functions introduced by Yedla \textit{et al.}, assuming that the potential function exists. We rewrite the density evolution of nonbinary SC-LDPC codes in an equivalent vector recursion form which is suited for the use of the potential function. We then discuss the existence of the potential function for the general case of vector recursions defined by multivariate polynomials, and give a method to construct it. We define a potential function in a slightly more general form than one by Yedla \textit{et al.}, in order to make the technique based on potential functions applicable to the case of nonbinary LDPC codes. We show that the potential function exists if a solution to a carefully designed system of linear equations exists. Furthermore, we show numerically the existence of a solution to the system of linear equations for a large number of nonbinary LDPC code ensembles, which allows us to define their potential function and thus prove threshold saturation. 

\end{abstract}

\section{Introduction}

Spatially-coupled low-density parity-check (SC-LDPC) codes have been shown to achieve outstanding performance for a myriad of channels and communication problems. Their excellent performance is due to the so-called \textit{threshold saturation phenomenon}. For the binary erasure channel (BEC) it was proved in \cite{KuRiUr11} that the belief propagation (BP) decoding of binary SC-LDPC codes \textit{saturates} to the maximum a posteriori (MAP) threshold of the underlying regular ensemble. This result was later extended to binary memoryless channels (BMS) \cite{KuMeRiUr10}, and the same threshold phenomenon has been observed for many channels and systems. Recently, an alternative proof technique for the threshold saturation phenomenon has been introduced in \cite{Yed14} and \cite{pfister-itw}, based on the notion of potential functions. The proof relies on the observation that a fixed point of the density evolution (DE) corresponds to a stationary point of the corresponding potential function. In \cite{Yed14}, for a class of coupled systems characterized by a scalar DE recursion, this technique was used to prove that the BP threshold saturates to the conjectured MAP threshold, known as the Maxwell threshold. This result was later extended in \cite{pfister-itw} to coupled systems characterized by vector DE recursions and, more recently, to SC-LDPC codes on BMS channels in \cite{macris-allerton}. It has also been shown that potential functions belong to a wider class of Lyapunov functions~\cite{SchBur2013}. 

Nonbinary LDPC codes designed over Galois fields of order $2^m$ (GF$(2^m)$), where $m$ is the number of bits per symbol, have received a significant interest in the last few years \cite{DaMa98,RaUr05}. For short-to-moderate block lengths, they have been shown to outperform binary LDPC codes. Nonbinary SC-LDPC codes have been considered recently in \cite{UcKaSa11} and \cite{PieGraCol13}. In \cite{PieGraCol13} it was shown that the MAP threshold of regular ensembles improves with $m$ and approaches the Shannon limit, and that, contrary to regular and irregular nonbinary LDPC codes for which the BP decoding threshold worsens for high values of $m$, the BP threshold of nonbinary SC-LDPC codes with large termination length improves with $m$ and tends to the Shannon limit. It was also empirically shown in \cite{PieGraCol13} that threshold saturation also occurs for nonbinary SC-LDPC codes.

One of the main contributions of this paper is to prove that, indeed, threshold saturation occurs for a large number of nonbinary SC-LDPC codes on the BEC. However, the contributions of this paper go further. We first prove the existence of a fixed point in the DE of nonbinary LDPC codes. To do so, we rewrite the DE in an equivalent vector recursion form based on complementary cumulative distribution function (CCDF) vectors, for which we can prove the monotonicity of the variable node and check node updates, and thus the existence of a fixed point in the DE. This equivalent form is also suited for the application of the proof technique introduced in \cite{Yed14,pfister-itw}, based on potential functions. However, here we use a more general definition of the potential function, since, as we will show, the potential function in the form defined in \cite{pfister-itw} does not exist for nonbinary LDPC codes. This motivated another main contribution of this paper: the analysis of the existence of the potential function for vector recursions defined by general multivariate polynomials \cite{pfister-itw,pfister-vector-itw}. We show that the potential function exists if a solution to a system of linear equations exists. As discussed in Section~\ref{Sec:DefD_F_G}, this system of linear equations is simply a new way of representing the design constraints from the definition of the potential function in \cite{pfister-itw}. 
Furthermore, our result is constructive: if a solution to the system of equations exists, then the potential function can be obtained by simply solving the system of equations. Throughout the paper we give some examples to compute the potential function.

The remainder of the paper is organized as follows. In Section~\ref{sec:DE}, we briefly discuss DE for nonbinary regular LDPC code ensembles and we present an equivalent formulation based on CCDF vectors. We also discuss the monotonicity of the variable node and check node updates and the existence of a fixed point in the DE. The potential function for the regular nonbinary LDPC code ensemble is discussed in Section~\ref{sec:PotentialRegular}, while in Section~\ref{Sec:ThresholdSaturation} we introduce the potential function for the spatially-coupled ensemble and give a proof of the threshold saturation. Some DE results are also provided in Section~\ref{Sec:ThresholdSaturation} for several SC-LDPC code ensembles. In Section~\ref{Sec:DefD_F_G}, we discuss the existence of the potential function and its calculation, and we provide some examples. Finally, some conclusions are provided in Section~\ref{sec:Conclusions}.

\subsection{Notation and Some Definitions}
\label{sec:notation}

We use upper case letters $F$ to denote scalar functions, bold lowercase letters $\xvec$ to denote vectors, and bold uppercase letters $\Xvec$ for matrices.
We denote by  $[\Xvec]_{i,j}$ the element in the $i$th row and $j$th column of a matrix $\Xvec$, and by $[\Xvec]_{i}$ the $i$th row of the matrix. Sometimes we will also use the alternative notation $x_{ij}=[\Xvec]_{i,j}$.
We assume all vectors to be row vectors, and we denote by ${\mathrm{vec}}( \Xvec)$ the row vector obtained by transposing the vector of stacked columns of matrix $\Xvec$. The transpose of a matrix/vector is denoted by $[\cdot]\T$. Let $\xvec\triangleq(x_1, \ldots, x_m)$ be a nonnegative vector of length $m$.  For two vectors $\xvec$ and $\yvec$ of length $m$, we use the partial order $\xvec  \preceq \yvec$ defined by $x_i \le y_i$, for $i=1,\ldots,m$.


The Jacobian of a scalar function $F(\xvec)$ is defined as
$$
F'=\derivF{F(\xvec)}{\xvec}\triangleq\left(\derivF{F}{x_1},\ldots,\derivF{F}{x_m}\right).
$$
Also, we define the Jacobian of a vector function $\fvec$ as
$$
\Fvec_{\mathrm d}(\xvec)=\fvec'(\xvec)\triangleq \left[\begin{array}{ccc}\derivF{f_1(\xvec)}{x_1} & \cdots & \derivF{f_1(\xvec)}{x_m} \\ \vdots & \ddots & \vdots \\ \derivF{f_m(\xvec)}{x_1} & \cdots & \derivF{f_m(\xvec)}{x_m}\end{array}\right] 
$$
where we denote by $f_k(\cdot)$ the $k$th component of the vector function $\fvec(\cdot)$. We also define the Hessian of a vector function $\fvec$ as
$$
\Fvec_{\mathrm{dd}}(\xvec) \triangleq\fvec''(\xvec).
$$

\label{sec:PotentialRegular}

\section{Density Evolution for $(\dl, \dr, m)$ and $(\dl, \dr, m, L, \dw)$ LDPC Code Ensembles over GF($2^m$)}
\label{sec:DE}
We consider transmission over the BEC with erasure probability $\varepsilon$, denoted as BEC($\varepsilon$), using nonbinary LDPC codes from an ensemble defined over the general linear group. The code symbols are elements of the binary vector space GF$(2^m)$, of dimension $m$, and we transmit on the BEC the $m$-tuples representing their binary images. We denote a regular nonbinary LDPC code ensemble over GF$(2^m)$ as  $(\dl, \dr, m)$, where $\dl$ and $\dr$ denote the variable node degree and the check node degree, respectively. Given a code in this ensemble, we associate 
to each edge of the corresponding bipartite graph a bijective linear mapping $\xi:\textrm{GF}(2^m) \rightarrow \textrm{GF}(2^m)$, chosen 
uniformly at random. The set of mappings is the general linear group $\textrm{GL}(2^m)$ over the binary field, 
which is the set of all $m\times m$ invertible matrices whose entries take values on $\{0,1\}$.
The design rate $r$ of a code in the ensemble does not depend on $m$ and can be expressed as 
$r=1-\frac{d_\texttt{v}}{d_\texttt{c}}$. We will also consider the regular $(\dl, \dr,m, L, \dw)$ SC-LDPC code ensembles, which are similar to the $(\dl, \dr, L, \dw)$ ensemble defined in \cite{KuRiUr11}, where $L$ denotes the spatial dimension and $\dw$ is the \textit{smoothing} parameter. This ensemble is obtained by placing $L$ sets of variable nodes of degree $\dl$ at positions $\{1,\ldots,L\}$. A variable node at position $t$ has $\dl$ connections to check nodes at positions in the range $\{t,t+1,\ldots,t+\dw-1\}$. For each connection, the position of the check node is uniformly 
and independently chosen from that range. 
A (terminated) $(\dl, \dr,m, L, \dw)$ SC-LDPC code ensemble is  defined by the parity-check matrix
\begin{equation*}
\label{Eq:Hmatrix}
\small
\H=\left[\begin{array}{c c c  }
 \H_0(1) & &   \\
 \vdots & \ddots &   \\
 \H_{\dw-1}(1) & &   \\
 & &     \H_0(L) \\
 &\ddots &  \vdots  \\
 & &   \H_{\dw-1}(L)
\end{array}\right].
\end{equation*} 
Each submatrix $\H_i(t)$ is a sparse $(M \dl/\dr)\times M$ nonbinary matrix, where $M$ is the number of variable nodes in each position and $M \dl/\dr$ is the number of check nodes in each position. It is important to note that the check node
degrees corresponding to the first and last couple
of positions is lower than $\dr$, i.e., the graph shows some irregularities. These irregularities lead to a locally better decoding (at the expense of a rate loss, which vanishes with $L$) and are the responsible for the outstanding performance of SC-LDPC codes.

In general, the messages exchanged in the BP decoding of nonbinary LDPC codes are real vectors $\boldsymbol{v}=(v_0,v_1,\ldots,v_{2^{m}-1})$, of length $2^m$, where $v_i$ represents the a posteriori probability that the code symbol $c$ is $c_i$. For instance, for $m=2$, there are four possible code symbols, $c_0=00$, $c_1=01$, $c_2=10$ and $c_3=11$, and the message $\vvec=(0.25,0.25,0.25,0.25)$ means that $\text{Pr}(c=00)=\text{Pr}(c=01)=\text{Pr}(c=10)=\text{Pr}(c=11)=0.25$. In the case of transmission over the BEC, the performance does not depend on the transmitted codeword and, without loss of generality, the transmission of the all-zero codeword can be considered \cite{RaUr05}. Under this assumption, the messages arising in the BP decoder assume a simplified form. In particular, the nonzero entries of a message $\vvec$ are all equal and the message itself is equivalent to a subspace of GF$(2^m)$. Since the nonzero elements of a message are equal, it is sufficient to keep track of the 
dimension of the messages~\cite{RaUr05}. 
We say that a message $\vvec$ has dimension $k$ if it has $2^k$ nonzero elements. For instance, the message $\vvec=(0.5,0.5,0,0)$ has dimension $1$. If a message coming from a node 
has dimension $k$, it means that the symbol is known to be one out of $2^k$ possible symbols (in this example, either $00$ or $01$) or, equivalently, that at that node $m-k$ 
relations on the bits composing the symbol are known. Consider as an example the three subspaces of dimension one of GF$(2^m)$, $S_1=\{00,01\}$, $S_2=\{00,10\}$ and $S_3=\{00,11\}$. 
Subspaces $S_1$ and $S_2$ are representative of the case where one bit has been recovered and the other is still erased, while $S_3$ represents the case 
where the two bits are erased but their sum modulo-2 is known. Therefore, the DE simplifies to the exchange of messages of length $m+1$, where the $i$th entry of the message is the probability that the message has dimension $i$. For more details the reader is referred to \cite{RaUr05}. In the following, we define the DE for nonbinary LDPC codes over the BEC for $(\dl, \dr, m)$ regular ensembles and  $(\dl, \dr,m, L, \dw)$ coupled ensembles.

\subsection{$(\dl, \dr, m)$ Regular LDPC Code Ensemble over GF($2^m$)}

Consider the $(\dl, \dr, m)$ ensemble over GF$(2^m)$, used for transmission over the BEC($\varepsilon$). Let $\xvecO^{(\ell)}\triangleq (x_{\circ 0}^{(\ell)}, \ldots, x_{\circ m}^{(\ell)})$ be the probability (row) vector of length $m+1$, where $x_{\circ i}^{(\ell)}$ is the probability that a message from variable nodes to check nodes at iteration ${\ell}$ has dimension $i$, $0 \le i \le m$. Likewise, $\yvecO^{(\ell)}\triangleq (y_{\circ 0}^{(\ell)}, \ldots, y_{\circ m}^{(\ell)})$ is the probability vector where $y_{\circ i}^{(\ell)}$ is the probability that a message from check nodes to variable nodes at iteration ${\ell}$ has dimension $i$. 

The variable node and check node DE updates at iteration $\ell$ are described by
\begin{align}
\xvecO^{(\ell)} = \fvecO(\yvecO^{(\ell)}; \varepsilon),~~~~
\yvecO^{(\ell)} = \gvecO (\xvecO^{(\ell-1)}) \nonumber
\end{align}
where $\fvecO=(f_{\circ 0},\ldots,f_{\circ m})$ and $\gvecO=(g_{\circ 0},\ldots,g_{\circ m})$ are, for a fixed $\varepsilon$, functions
from $[0,1]^{m+1}$ to $[0,1]^{m+1}$, defined by
\begin{align}
\label{eq:bij1}
\fvecO(\yvecO; \varepsilon)  &\triangleq \pvecO(\varepsilon) \boxdot \left( \boxdot^{\dl-1} \yvecO \right)
\\
\label{eq:bij2}
\gvecO(\xvecO) & \triangleq \boxtimes^{\dr-1} \xvecO 
\end{align}
where we define $\boxdot^{\dl-1} \avecc = \avecc \boxdot \avecc \boxdot \cdots \boxdot \avecc$ with $\dl-1$ terms $\avecc$ (i.e., $\boxdot^1 \avecc = \avecc$), and
$\boxtimes^{\dr-1} \avec = \avecc \boxtimes \avecc \boxtimes \cdots \boxtimes \avecc$ with $\dr-1$ terms $\avecc$ (i.e., $\boxtimes^1 \avecc = \avecc$). $\pvecO$ is a row vector of length $m+1$, the $i$th element of which being the probability that the channel message has dimension $i$,
\begin{equation}
\label{eq:pvecO}
{\pvecO}_{i}(\varepsilon) \triangleq \binom{m}{i} \varepsilon^i (1-\varepsilon)^{m-i} , \quad i=0,\cdots, m\, .
\end{equation}


For two probability vectors $\avecc$ and $\bvecc$ of length $m+1$, the operations $\avecc \boxdot \bvecc$ and $\avecc \boxtimes \bvecc$ are defined as
\begin{align} 
[\avecc \boxdot \bvecc]_k&\triangleq\sum_{i=k}^m \sum_{j=k}^{m+k-i} V^m_{i,j,k} a_{\circ i} b_{\circ j}, \quad k=0,\ldots,m \label{eq:boxdot}\\
[\avecc \boxtimes \bvecc]_k&\triangleq \sum_{i=0}^k \sum_{j=k-i}^{k} C^m_{i,j,k} a_{\circ i} b_{\circ j}, \quad k=0,\ldots,m \label{eq:boxtimes}
\end{align}
where $V^m_{i,j,k}$ is the probability of choosing a subspace of dimension $j$ whose intersection with a subspace of dimension $i$ has dimension $k$, and $C^m_{i,j,k}$ is the probability of choosing a subspace of dimension $j$ whose sum with a subspace of dimension $i$ has dimension $k$,
\begin{align}
V^m_{i,j,k}&=\frac{{\GaussBin}_{i,k} {\GaussBin}_{m-i,j-k} 
2^{(i-k)(j-k)}}{{\GaussBin}_{m,j}} \nonumber\\
C^m_{i,j,k}&=\frac{{\GaussBin}_{m-i,m-k} {\GaussBin}_{i,k-j} 
2^{(k-i)(k-j)}}{{\GaussBin}_{m,m-j}}.\nonumber
\end{align}
${\GaussBin_{m,k}}$ is the Gaussian binomial coefficient,
\begin{equation}\label{e:gauss_bin_coeff}
{\GaussBin}_{m,k}=\left[\!\!\! \begin{array}{c} m\\ k \end{array} \!\!\!\right]= 
\begin{cases}
1, & \text{if } k=m \text{ or } k=0\\
\displaystyle \prod_{\ell=0}^{k-1}\frac{2^m-2^\ell}{2^k-2^\ell}, & \text{if } 0< k< m\\
0, & \text{otherwise}
\end{cases}
\end{equation}
which gives the number of different subspaces of dimension $k$ of GF$(2^m)$.

In Appendix~\ref{app:UsefulResults} we derive some results on the coefficients $V^m_{i,j,k}$ and $G_{m,k}$ which will be useful for the proof of Theorem~\ref{The:Monotonicity} below.

The DE recursion can be written in compact form as
\begin{equation}
\label{eq:DElrm0}
\xvecO^{(\ell+1)} = \fvecO(\gvecO(\xvecO^{(\ell)});\varepsilon)
\end{equation}
and starts from $\xvecO^{(0)}=\pvecO$. Note that decoding is successful when the DE converges to $\xvecO^{(\infty)}=(1,0,\ldots,0)$. The DE equations (\ref{eq:bij1}) and (\ref{eq:bij2}) were first defined in \cite{RaUr05}.

In the following, we rewrite the DE recursion in (\ref{eq:DElrm0}) in a more suitable form to prove threshold saturation based on potential functions. The reason for this is that the approach in \cite{pfister-itw} requires monotone vector functions for the variable node and check node updates. It can be shown that $\fvecO(\yvecO;\eps)$ and $\gvecO(\xvecO)$ are not monotone, and, therefore, cannot be used directly.

We introduce the notion of a CCDF vector.

\begin{definition}
Given a probability vector $\xvecO=(x_{{\circ 0}},x_{{\circ 1}},\ldots,x_{{\circ m}})$ of length $m+1$, we define a CCDF vector $\xvec=(x_1,\ldots,x_m)$ of length $m$ by means of the following bijective function ${\cal H}$,
\begin{align}
\label{eq:Map1}
\xvecO \stackrel{\cal H}{\rightarrow} \xvec & : \quad x_i=\sum_{k=i}^m x_{\circ k}, \quad \quad ~1\le i \le m \\
\label{eq:Map2}
\xvec \stackrel{{\cal H}^{-1}}{\rightarrow} \xvecO & :  \quad x_{\circ i}=\begin{cases} 1-x_1,& i=0\\ x_i-x_{i+1},&1 \le i < m\\
x_m,&i=m. \end{cases}
\end{align}
\end{definition}


%
%

By considering the CCDF vectors $\xvec=(x_1,\ldots,x_m)$, $\yvec=(y_1,\ldots,y_m)$ and $\pvec=(p_1,\ldots,p_m)$ and the mapping $\mathcal{H}$ above, we can define new vector functions $\fvec(\yvec;\varepsilon)=(f_1,\ldots,f_m)=\mathcal{H}(\fvecO)$ and $\gvec(\xvec)=(g_1,\ldots,g_m)=\mathcal{H}(\gvecO)$, with
\begin{align}
\label{eq:fi}
f_i(\yvec;\varepsilon) &=\sum_{k=i}^m f_{\circ k}(\yvecO;\varepsilon)
\end{align}
and
\begin{align*}
g_i(\xvec)&=\sum_{k=i}^m g_{\circ k}(\xvecO).
\end{align*}

Then, the variable node and check node DE updates at iteration $\ell$ can be written in an equivalent form as
\begin{equation}
\label{eq:DEupdatesEq}
\xvec^{(\ell)} = \fvec(\yvec^{(\ell)}; \varepsilon),~~~~
\yvec^{(\ell)} = \gvec (\xvec^{(\ell-1)})
\end{equation}
and the DE recursion (\ref{eq:DElrm0}) can be rewritten as
\begin{equation}
\label{eq:DElrm}
\xvec^{(\ell+1)} = \fvec(\gvec(\xvec^{(\ell)});\varepsilon).
\end{equation}
\begin{theorem}
\label{The:Monotonicity}
The functions $\fvec(\xvec; \varepsilon)$ and $\gvec(\xvec)$ are 
increasing in $\xvec$ with respect to the partial order $\preceq$.
\end{theorem}
\begin{IEEEproof}
The proof is given in Appendix~\ref{app:ProofTh1}.
\end{IEEEproof}

For later use, we denote by $\cal X$ the set of all possible values of $\xvec$. Likewise, we denote by $\cal Y$ and $\cal E$ the set of all possible values of $\yvec$ and $\varepsilon$, respectively. 
Therefore,  
\begin{align}
{\cal E}&: \quad 0 \le \varepsilon \le 1\nonumber\\
{\cal X}&: \quad 0 \le x_i \le 1, &\ 1 \le i\le m\nonumber\\
{\cal Y}&: \quad 0 \le y_i \le 1, &\ 1 \le i\le m\nonumber
\end{align}

\begin{definition}
Let $\xvec\in\mathcal{X}$, $\varepsilon\in\mathcal{E}$, and $\xvec^{(0)}=\xvec$. Then $\xvec^\infty(\xvec;\varepsilon)\triangleq \lim_{\ell\rightarrow\infty}\fvec(\gvec(\xvec^{(\ell)};\varepsilon))$.
\end{definition}

\begin{corollary}
For regular nonbinary LDPC codes,  $\xvec^\infty(\onevec;\varepsilon)$  exists, i.e., the DE converges to a fixed point $\xvec^\infty(\onevec;\varepsilon)$. 
\end{corollary}
\begin{IEEEproof}
Since $\fvec(\xvec; \varepsilon)$ and $\gvec(\xvec)$ are increasing in $\xvec$ with respect to the partial order $\preceq$, the DE sequence in (\ref{eq:DEupdatesEq}) is monotonic with the number of iterations and, thus, converges to a limit. This limit is a fixed point of the recursion because the function $\fvec ( \gvec (\xvec);\varepsilon)$ is continuous in $\xvec$.
\end{IEEEproof}

Note that successful decoding corresponds to convergence of the DE equation (\ref{eq:DElrm}) to the fixed point $\xvec^{(\infty)}={\mathbf 0}=(0,0,\ldots,0)$, because the CCDF of $(1,0,\ldots,0)$ is $(0,\ldots,0)$.


For nonbinary codes and for some $\varepsilon$, the domain and range of $\fvec(\yvec;\eps)$ and $\gvec(\xvec)$ are given by $\cal X$ and $\cal Y$ respectively.
Also, the vector functions $\fvec(\yvec;\eps)$ and $\gvec(\xvec)$ have several properties which will be useful for the proof of threshold saturation in Sections~\ref{sec:PotentialRegular} and \ref{Sec:ThresholdSaturation}.
{\lemma\label{lem:Lemma1}
Consider 
$\fvec(\yvec; \varepsilon)$ 
and $\gvec(\xvec)$ 
defined above. For $\xvec\in{\cal X}$ and $\yvec\in{\cal Y}$,
\begin{enumerate}
\item $\fvec(\yvec; \varepsilon)$ and $\gvec(\xvec)$ are nonnegative vectors;
\item $\fvec(\yvec; \varepsilon)$ is differentiable in $\yvec$ and $\gvec(\xvec)$ is twice differentiable in $\xvec$;
\item $\fvec(\zerovec; \varepsilon) = \fvec (\yvec; 0)=\gvec(\zerovec)=\zerovec$;
\item 
$\Gvec_{\mathrm d}(\xvec)> 0$, 
and it is positive definite for $\xvec\in{\cal X}\backslash\{\zerovec\}$;
\item $\fvec(\yvec; \varepsilon)$ is strictly increasing with $\varepsilon$.
\end{enumerate}
}

\begin{IEEEproof}
The first property follows from the fact that $\fvec(\yvec;\eps)$ and $\gvec(\xvec)$ are CCDF vectors. The second property follows from $\fvec(\yvec;\eps)$ and $\gvec(\xvec)$ being multivariate polynomials. The third property follows from $(1,0,\ldots,0)\boxtimes (1,0,\ldots,0)= (1,0,\ldots,0)$, $(1,0,\ldots,0)\boxdot \xvecO = (1,0,\ldots,0)$ and $\pvecO=(1,0,\ldots,0)$ for $\varepsilon=0$. For the fourth property,
due to the fact that the coefficients $C^m_{i,j,k} =0$ for $k < i,j$, $\Gvec_{\mathrm d}(\xvec)$ is of special form. In fact, it is a lower triangular matrix, whose entries $(i,j)$, $i \le j$, are multivariate polynomials in $\xvec$ with positive coefficients. Thus, for any $\xvec>\zerovec$, the lower triangular part of $\Gvec_{\mathrm d}(\xvec)$ is positive. From the positiveness of the elements of the diagonal, it follows that all eigenvalues of $\Gvec_{\mathrm d}(\xvec)$ are positive for $\xvec \in{\cal X}\backslash\{\zerovec\}$ and, thus, $\Gvec_{\mathrm d}(\xvec)$ is positive definite\footnote{In \cite{pfister-itw}, the positive definite property is needed for matching the potential threshold and the Maxwell threshold. Furthermore, a positive definite matrix is invertible, which will be used in Lemma~\ref{Lem:Potential}.}.
Finally, to prove the fifth property we can write
\begin{equation}
\label{eq:derivative_f_eps}
\derivF{f_i}{\varepsilon}=\sum_{k=1}^m\derivF{f_i}{p_k}\derivF{p_k}{\varepsilon}.
\end{equation}
As shown in Appendix~\ref{app:ProofTh1}, $\derivF{f_i}{p_k}>0$. The second term in each product is
\begin{align}
\label{eq:PartialDer}
\derivF{p_k}{\varepsilon}&=\sum_{\ell=0}^m{m \choose \ell}\varepsilon^{\ell-1}(1-\varepsilon)^{m-\ell-1}(\ell-\varepsilon m)\mathbbm{1}{ \{\ell \ge k\} }\\
&=\frac{1}{(1-\varepsilon)\varepsilon}\sum_{\ell=0}^m {m \choose \ell}\varepsilon^{\ell}(1-\varepsilon)^{m-\ell}(\ell-\varepsilon m)\mathbbm{1}{ \{\ell \ge k\} }.
\end{align}
It is easy to verify that
\begin{align}
\label{eq:PartialDer3}
\derivF{p_k}{\varepsilon}&>\derivF{p_0}{\varepsilon}=0,&\forall k \ge 1 \text{ and } 0<\varepsilon<1.
\end{align}
It then follows that (\ref{eq:derivative_f_eps}) is positive for all values of $i$, \\ $i=1,\ldots, m$, therefore $\fvec(\yvec;\eps)$ is increasing in $\varepsilon$.
\end{IEEEproof}

\subsection{$(\dl, \dr,m, L, \dw)$ SC-LDPC Code Ensemble over GF($2^m$)}

Consider the $(\dl, \dr,m, L, \dw)$ ensemble over GF$(2^m)$ and transmission over the BEC($\varepsilon$).  
In the form of (\ref{eq:DElrm}), the DE equations for the $(\dl, \dr,m, L, \dw)$ ensemble can be written as 
\begin{align}
\xvec_{i} = \frac{1}{\dw} \sum_{k=0}^{\dw-1} \fvec(\yvec_{i-k};\varepsilon_{i-k}),~~~~
\yvec_{i}= \frac{1}{\dw} \sum_{k=0}^{\dw-1}\gvec(\xvec_{i+k})\nonumber
\end{align} 
where $1 \le i <L+\dw$, and 
$$\varepsilon_i = \begin{cases} \varepsilon,& 1 \le i \le L\\ 0,&1 \le i-L <\dw. \end{cases}
$$

Collect the CCDF vectors $\xvec_i$ into the $(L +\dw-1)\times m$ matrix 
$\Xvec = (\xvec_1\T, \ldots, \xvec_{L+\dw-1}\T)\T$ and the CCDF vectors $\yvec_i$ into the $L\times m$ matrix $\Yvec = (\yvec_1\T, \ldots, \yvec_{L}\T)\T$.
Also, let $\Avec$ be the $L \times (L+\dw-1)$ matrix
\vspace{-1.1cm}
 \[ \vphantom{
    \begin{matrix}
    \overbrace{XYZ}^{\mbox{$R$}}\\ \\ \\ \\ \\ \\ 
    \underbrace{pqr}_{\mbox{$S$}}
    \end{matrix}}%
\begin{matrix}
\vphantom{a}\\ 
\end{matrix}%
\Avec=\frac{1}{\dw}\begin{pmatrix}
1 & 1 & \cdots & 1 & 0 & 0 & \cdots & 0\\
 0&1&\cdots &    1& 1& 0& \cdots &0\\
 \vdots&\vdots&\ddots &    \vdots& \vdots& \vdots& \ddots &\vdots\\
\coolunder{\dw}{0 & 0 & 0 & 0 &} \coolunder{L-1}{0 & 1 & \cdots & 1}\\
\end{pmatrix}.%
\begin{matrix}
\end{matrix}\]
\vspace{-0.7cm}
 
The fixed-point DE equation for the $(\dl, \dr,m, L, \dw)$ ensemble can then be written in matrix form, similarly as in \cite{pfister-itw}
$$\Xvec = \Avec\T \Fvec (\Avec \Gvec(\Xvec); \varepsilon)$$
where $\Fvec(\Yvec;\varepsilon)$ is an $L\times m$ matrix, $\Fvec(\Yvec;\varepsilon) = (\fvec(\yvec_1;\varepsilon)\T, \ldots, \fvec(\yvec_L;\varepsilon)\T )\T$, and $\Gvec(\Xvec)$ is an {$(L+\dw-1)\times m$ matrix, $\Gvec(\Xvec) = (\gvec(\xvec_1)\T, \ldots, \gvec(\xvec_{L+\dw-1})\T )\T$. }


\section{Potential Function for the $(\dl,\dr,m)$ Ensemble}
\label{sec:PotentialRegular}
\label{Sec:PotentialFunction}
The DE equation (\ref{eq:DElrm}) for the $(\dl,\dr,m)$ regular ensemble describes a vector system for which we can properly define a potential function, similarly to \cite{pfister-itw}.
{\definition
\label{def:PotentialFuntion}
The potential function $U(\xvec; \varepsilon)$ of the system defined by functions $\fvec(\yvec;\eps)$ and $\gvec(\xvec)$ above  is given by
\begin{align}
\label{eq:PotentialFunction}
U(\xvec; \varepsilon) 
\triangleq \gvec(\xvec)\Dvec \xvec\T -G(\xvec)-F(\gvec(\xvec);\varepsilon)
\end{align}
where $F: {\cal X} \times {\cal E} \mapsto {\mathbb R}$ and $G: {\cal Y} \times {\cal E} \mapsto {\mathbb R}$ are scalar functions that satisfy
$F(\zerovec)=0$, $G(\zerovec)=0$, $F'(\yvec; \varepsilon) = \fvec(\yvec;\varepsilon) \Dvec$, and
$G'(\xvec) = \gvec(\xvec) \Dvec$,
for a symmetric, invertible $m\times m$ matrix $\Dvec$ with positive elements $d_{ij}$.}

The definition of $U(\xvec; \varepsilon)$ above is slightly more general than the one in \cite{pfister-itw}, since $\Dvec$ is assumed to be a positive, symmetric and invertible matrix,  instead of being a diagonal matrix as in \cite{pfister-itw}. In Section~\ref{Sec:DefD_F_G}, we discuss the existence of the potential function for the general case of vector recursions defined by multivariate polynomials, and we show that, for the nonbinary codes considered here, the potential function in the form of (\ref{eq:PotentialFunction}), with $F'(\yvec; \varepsilon) = \fvec(\yvec;\varepsilon) \Dvec$, and $G'(\xvec) = \gvec(\xvec) \Dvec$, does not exist for a diagonal matrix $\Dvec$. The fact that we assume $\Dvec$ in Definition~\ref{def:PotentialFuntion} to be positive, symmetric and invertible is used to prove Assertion 1 and Assertion 2 in Lemma~\ref{Lem:Potential} below. For a positive, symmetric and invertible matrix $\Dvec$, it is shown in Section~\ref{sec:F-and-G-nb} that the potential function may exist.




{
{\definition 
For $\xvec \in {\cal X}$ and $\varepsilon \in {\cal E}$,
$\xvec$ is a fixed point of the DE if $\xvec = \fvec(\gvec(\xvec);\varepsilon)$; 
$\xvec$ is a stationary point of the potential function if $U'(\xvec;\varepsilon)=\zerovec$.
}

Let the fixed point set be defined as $${\cal F} \triangleq \{(\xvec;\varepsilon) | \xvec=\fvec(\gvec(\xvec);\varepsilon)\}.$$

{\lemma
\label{Lem:Potential}
For the vector system defined by $\fvec(\yvec;\eps)$ and $\gvec(\xvec)$, the following assertions hold.
\begin{enumerate}
\item $\xvec \in {\cal X}$ is a fixed point if and only if it is a stationary point of the potential function $U(\xvec;\varepsilon)$;
\item $U(\xvec;\varepsilon)$ is strictly decreasing in $\varepsilon$, for $\xvec \in {\cal X} \backslash \zerovec$ and $\varepsilon \in \cal{E}$;
\item $U'(\xvec; \varepsilon)$ is strictly decreasing in $\varepsilon$;
\item For some $\varepsilon_1 > 0$ and $\varepsilon_2 > 0$ such that $\varepsilon_1 \neq \varepsilon_2$, if $(\xvec_1, \varepsilon_1) \in {\cal F}$ and $(\xvec_2, \varepsilon_2) \in {\cal F}$, then $\xvec_1\not=\xvec_2$.
\end{enumerate}
}
\begin{IEEEproof}
\begin{enumerate}
\item $U'(\xvec;\varepsilon)$ is obtained as (see Appendix~\ref{app:DerivativeU})
\begin{equation}
\label{eq:DerU}
U'(\xvec;\varepsilon) = (\xvec-\fvec(\gvec(\xvec);\varepsilon))\Dvec\gvec'(\xvec).
\end{equation}
Since $\Dvec$ is positive, it follows $\Dvec\gvec'(\xvec)>0$. Furthermore, since $\Dvec$ is invertible, $\Dvec \gvec'(\xvec)$ is also invertible. Therefore, if $\xvec$ is a stationary point of $U(\xvec;\varepsilon)$, it follows $\xvec-\fvec(\gvec(\xvec);\varepsilon)=0$, i.e., $\xvec$ is a fixed point of the DE. The converse statement is trivial.
\item $U(\xvec;\varepsilon)$ is given in (\ref{eq:PotentialFunction}). The only term depending on $\varepsilon$ is $F(\yvec;\varepsilon)$. Therefore, it is sufficient to prove that $F(\yvec;\varepsilon)$ is increasing in $\varepsilon$. $F(\yvec;\varepsilon)$ is the line integral of $\fvec(\yvec;\varepsilon)\Dvec$,
\begin{align*}
F(\yvec;\varepsilon)=\int_{\zerovec}^{\yvec}\fvec(\zvec;\varepsilon)\Dvec d\zvec .
\end{align*}
Since the integrand is an increasing function of $\varepsilon$ (which follows from the fact that $\fvec(\yvec;\varepsilon)$ is increasing in $\eps$ (see Lemma~\ref{lem:Lemma1}) and $\Dvec$ is a positive matrix), then the line integral is also an increasing function of $\varepsilon$.

\item $U'(\xvec;\varepsilon)$ is given in (\ref{eq:DerU}). Note that the only term that depends on $\varepsilon$ is $\fvec(\gvec(\xvec);\varepsilon)$, therefore it is sufficient to show that $\fvec(\gvec(\xvec);\varepsilon)$ is increasing in $\varepsilon$. This was already proven in Lemma~\ref{lem:Lemma1}.
\item The fourth assertion is true because $(\xvec - \fvec(\gvec(\xvec);\varepsilon))$ is strictly decreasing in $\varepsilon$.


%
\end{enumerate}
\end{IEEEproof}

We can now define the BP and the potential thresholds, denoted respectively by $\thBP$ and $\varepsilon^*$.
{\definition
The BP threshold is
$$\thBP \triangleq \sup \left\{ \varepsilon \in {\cal E} | \xvec^\infty(\onevec;\varepsilon)=\zerovec \right\}.$$
}

In order to define the potential threshold $\varepsilon^*$, let us define the \textit{energy gap} $\Egap (\varepsilon)$ with the help of the following definition of the basin of attraction of the fixed point $\xvec^{(\infty)}=\zerovec$ (successful decoding) \cite{pfister-itw}:
{\definition
The basin of attraction for $\xvec^{(\infty)} = \zerovec$
is
$${\cal U}_{\zerovec}(\varepsilon) \triangleq \{ \xvec \in {\cal X} | \xvec^{\infty}(\xvec;\varepsilon) = \zerovec \}.$$
}
{\definition The energy gap $\Egap(\varepsilon)$ for some $\varepsilon$, $\thBP \le \varepsilon \le \varepsilon^*$, is defined as
$$\Egap(\varepsilon) \triangleq \inf_{\xvec \in {\cal X}\backslash {\cal U}_{\zerovec} (\varepsilon)} U(\xvec; \varepsilon).$$
}
We are ready to define $\varepsilon^*$.

{\definition
The potential threshold is
$$\varepsilon^* \triangleq \sup \left\{ \varepsilon \in (\thBP,1] \   | \  \Egap (\varepsilon) > 0, \ \forall \xvec \in {\cal X} \right\}.$$


{\remark
The definition of $\varepsilon^*$ is similar to the one given in \cite{pfister-vector-itw}. It is equivalent to the definition given in \cite{pfister-itw} if $U(\xvec; \varepsilon)$ is positive for $\varepsilon \in (\thBP, \varepsilon^*)$ and $\Egap (\varepsilon^*) = 0$.
}

{\remark

It has been shown for several systems in \cite{Yed14, pfister-vector-itw, pfister-vector-arxiv}, that the MAP threshold $\thMAP$ and the potential threshold $\varepsilon^*$ are identical. The idea to prove this result is given in \cite{pfister-vector-arxiv}. It implies the calculation of the trial entropy $P(x)$, a quantity related to the BP EXIT function $h^{\mathrm{BP}}(\varepsilon)$, which was first defined in \cite{measson2008}. Unfortunately, this approach cannot be applied to nonbinary LDPC codes, since the general expression of $h^{\mathrm{BP}}(\varepsilon)$ for an arbitrary value of $m$ is not yet known. In this paper, we do not address the question of the equality between $\thMAP$ and $\varepsilon^*$ for nonbinary LDPC codes.
}

}


\section{Potential Function for the Spatially-Coupled System and a Proof of Threshold Saturation}
\label{Sec:ThresholdSaturation}
{
{\definition 
\label{def:PotentialFunctionVec}
The potential function $U(\Xvec;\varepsilon)$ for the spatially-coupled case is defined similarly as in \cite{pfister-itw}
\begin{equation}
\label{eq:PotFunctionVect}
U(\Xvec;\varepsilon) \triangleq \tr(\Gvec(\Xvec) \Dvec \Xvec\T ) - G(\Xvec) - F(\Avec \Gvec(\Xvec);\varepsilon)
\end{equation}
where  
{
$G'(\Xvec) = \sum_{i=1}^{L+w-1} G'(\xvec_i) = \sum_{i=1}^{L+w-1} \gvec(\xvec_i) \Dvec$,}
and
{
$F'(\Xvec) = \sum_{i=1}^L F'(\xvec_i) = \sum_{i=1}^L \fvec(\xvec_i) \Dvec$.}

}


To prove threshold saturation, we will need the partial derivative of $U(\Xvec;\varepsilon)$. It is given in the following theorem.
{\theorem \label{Th:Uprime}
The partial derivative of $U(\Xvec;\varepsilon)$ is 
\begin{equation}\nonumber
U'(\Xvec;\varepsilon)\triangleq \left[\begin{array}{c}{\deriv {{\xvec}_1}} U(\Xvec;\varepsilon)  \\ \vdots \\ {\deriv {{\xvec}_L}} U(\Xvec;\varepsilon) \end{array}\right]
\end{equation}
}
where the $i$th row of $U'(\Xvec;\varepsilon)$ is 
\begin{equation}\nonumber
[U'(\Xvec;\varepsilon)]_i=\left({\xvec}_i-[\Avec\T]_i \Fvec(\Avec\Gvec(\Xvec);\varepsilon)\right)\Dvec\Gvec_{\mathrm d}(\xvec_i).
\end{equation}
\begin{IEEEproof}
The proof of the theorem is given in Appendix~\ref{app:ProofTh2}.
\end{IEEEproof}

We also need the following property of $U''(\Xvec; \varepsilon)$.
{\lemma 
The norm of the second derivative of $U(\Xvec;\varepsilon)$ is upper bounded by
$$||U''(\Xvec;\varepsilon)||_{\infty} \le ||\Dvec ||_{\infty} (\alpha + \beta + \alpha^2 \gamma) = K$$
where 
$\alpha= \sup_{\xvec \in {\cal X}} ||\Gvec_{\mathrm d}(\xvec)||_{\infty}$,
$\beta= \sup_{\xvec \in {\cal X}} || \Gvec_{\mathrm{dd}}(\xvec)||_{\infty}$,
and 
$\gamma= \sup_{\yvec \in {\cal Y}} ||\Fvec_{\mathrm d}(\yvec;\varepsilon)||_{\infty}$.
}
}
\begin{IEEEproof}
The proof of the lemma is similar to the one given in \cite[Lemma 8]{pfister-itw}.
\end{IEEEproof}

The following theorem proves successful decoding for $\varepsilon<\varepsilon^*$, i.e., the BP decoder saturates to the potential threshold for large enough values of $\dw$.
{\theorem Given the spatially coupled $(\dl, \dr,m,L,\dw)$ LDPC code ensemble, 
for $\varepsilon<\varepsilon^*$ and $\dw> \frac{m K}{2 \Egap(\varepsilon)}$, the only fixed point of the system is $\xvec^{\infty}=\zerovec$.
}
\begin{IEEEproof}
The proof of the theorem follows the same lines as the proof in \cite{pfister-vector-itw} and \cite{pfister-vector-arxiv} and is omitted for brevity.
\end{IEEEproof}

\subsection{Numerical Results}

\begin{table}[!t]
\addtolength{\tabcolsep}{-0.4mm}
\caption{DE thresholds for nonbinary SC-LDPC codes}
\vspace{-2ex}
\begin{center}\begin{tabular}{cccccccc}
\toprule
Ensemble& Rate  & $\varepsilon^1_{\text{BP}}$ & $\varepsilon^3_{\text{BP}}$ & $\varepsilon^5_{\text{BP}}$ & $\varepsilon^8_{\text{BP}}$ & $\varepsilon_{\text{MAP}}$ & $\delta_{\rm{Sh}}$\\
\otoprule
$(3,6)$   & $1/2$  &0.4880& 0.4978  & 0.4995 & 0.4998 & 0.4999 & 0.0002\\[0.5mm]
$(3,9)$   & $2/3$  &0.3196 & 0.3307  & 0.3328 & 0.3331 & 0.3332 & 0.0002\\[0.5mm]
$(3,12)$  & $3/4$ &0.2372 & 0.2476 & 0.2495 & 0.2497 &0.2499 & 0.0003\\[0.5mm]
$(3,15)$  & $4/5$ &0.1886 & 0.1978 & 0.1995 & 0.1996 & 0.1999 & 0.0004\\[0.5mm]
\bottomrule
\end{tabular} \end{center}
\label{Tab:DEThreshold} 
\end{table}

In Table~\ref{Tab:DEThreshold} we give the BP threshold of nonbinary SC-LDPC code ensembles with $\dl=3$ for several code rates and $m=1,3,5$ and $8$, denoted by $\varepsilon^1_{\text{BP}}$, $\varepsilon^3_{\text{BP}}$, $\varepsilon^5_{\text{BP}}$, and $\varepsilon^8_{\text{BP}}$, respectively, for $L\rightarrow\infty$. In particular, we consider the $(\dl, \dr, m, L)$ coupled ensemble defined in~\cite{KuRiUr11}, properly extended to the nonbinary case. We observe that, for a given rate, the BP threshold improves with increasing values of $m$. A significant improvement is observed from $m=1$ (binary) to $m=3$. It is interesting to note that $\varepsilon_{\text{BP}}$ approaches the Shannon limit as $m$ increases (the last column of the table gives the gap to the Shannon limit for the coupled ensembles with $m=8$, $\delta_{\rm{Sh}}$). We have observed that the BP threshold tends to the MAP threshold $\varepsilon_{\text{MAP}}$ for large $L$ for all values of $m$, suggesting that threshold saturation to the MAP threshold occurs. As an example, we report in the table the MAP threshold for $m=8$.



\section{Existence of the Potential Function: Calculation of $F(\yvec; \varepsilon)$ and $G(\xvec)$, and \\Properties of $\Dvec$}
\label{Sec:DefD_F_G}

In this section, we discuss the existence of the potential function in the form of (\ref{eq:PotentialFunction}) and its calculation. The existence of $U(\xvec;\epsilon)$ in (\ref{eq:PotentialFunction}) depends crucially on the existence of the functions $F(\yvec;\epsilon)$ and $G(\xvec)$ that satisfy
\begin{align}
\label{eq:conditionsa}
F'(\yvec;\varepsilon)=\fvec(\yvec;\varepsilon)  \Dvec,\quad\quad\quad
G'(\xvec)=\gvec(\xvec)  \Dvec .
\end{align}

We first discuss (Section~\ref{sec:F-and-G} below) the calculation of $F(\yvec; \varepsilon)$ and $G(\xvec)$ in the general case where $\fvec(\yvec;\eps)$ and $\gvec(\xvec)$ are defined by multivariate polynomials \cite{pfister-itw,pfister-vector-itw}, without making any assumption on the form of matrix $\Dvec$. The discussion encompasses any class of sparse-graph codes (coupled or not) used for transmission over the BEC. In particular, we show that the potential function exists if there exists a solution to a carefully defined system of linear equations. In this case, $F(\yvec; \varepsilon)$ and $G(\xvec)$ can be obtained by solving the system of linear equations. We then consider in Section~\ref{sec:ChoiceD} the choice of $\Dvec$ so that a solution to the system of linear equations exists. The particular case of nonbinary $(\dl,\dr,m)$ and $(\dl,\dr,m, L, \dw)$ LDPC ensembles is considered in Section \ref{sec:F-and-G-nb}.


\subsection{Calculation of $F(\yvec; \varepsilon)$ and $G(\xvec)$ from $\fvec(\yvec; \varepsilon)$ and $\gvec(\xvec)$}
\label{sec:F-and-G}

The problem of calculating $F(\yvec; \varepsilon)$ and $G(\xvec)$ corresponds to the problem of reconstructing two multivariate polynomial functions from their multivariate polynomial gradient vector functions $\fvec(\yvec; \varepsilon)$ and $\gvec(\xvec)$. 
The main result of this section (Theorem \ref{theorem:existence-F-G}) is that this reconstruction problem is equivalent to the relatively simple problem of solving a system of linear equations. 
It is important to note that, indeed, (\ref{eq:conditionsa}) leads to a system of linear equations. Therefore, to determine the existence of $F(\yvec; \varepsilon)$ and $G(\xvec)$ one needs to determine whether a solution of this system exists or not.  

First, note that (\ref{eq:conditionsa}) can be equivalently written as
\begin{align}
\label{eq:F1}
\derivF{F}{y_i}(\yvec;\varepsilon)&=\sum_{j=1}^m d_{ji} f_j(\yvec, \varepsilon)\\
\label{eq:G1}
\derivF{G}{x_i}(\xvec)&=\sum_{j=1}^m d_{ji} g_j(\xvec)
\end{align}
for $i=1,\ldots,m$. 

In what follows, we will make use the following definition.
\begin{definition}
Let $\setf{j}$ and $\setg{j}$, $j=1\ldots,m$, be the set of nonzero coefficients of the multivariate polynomial functions $f_j(\yvec;\varepsilon)$ and $g_j(\xvec)$, respectively, defined as
\begin{align}
\label{eq:coefff}
\setf{j}&\triangleq\left\{(i_1,\ldots,i_m):\mathrm{coeff}\left(f_j(\yvec),\prod_{k=1}^m y_k^{i_k}\right)\neq 0\right\}\\
\label{eq:coeffg}
\setg{j}&\triangleq\left\{(i_1,\ldots,i_m):\mathrm{coeff}\left(g_j(\xvec),\prod_{k=1}^m x_k^{i_k}\right)\neq 0\right\}.
\end{align}
Let also $\set{f}{} = \cup_j \set{f}{j}$ and $\set{g}{} = \cup_j \set{g}{j}$.
\end{definition}

Now, in order that (\ref{eq:F1})--(\ref{eq:G1}) hold, the sets of coefficients of the monomials in the left and right hand side of (\ref{eq:F1})--(\ref{eq:G1}) must be the same. We define in the following the sets of coefficients which are related to the left hand side of (\ref{eq:F1})--(\ref{eq:G1}).

\begin{figure*}
{\footnotesize
\begin{align}
\label{eq:Fjk-def}
\set{F}{j,k}&\triangleq \left\{(i_1,\ldots,i_m):  \mathrm{coeff}\left(\int_0^{y_k} f_j(y_1,\ldots,y_{k-1},z,y_{k+1},\ldots y_m)dz,\prod_{l=1}^m y_l^{i_l}\right)\neq 0\right\}, \quad \text{for }j,k=1,\ldots,m\\
\label{eq:Gjk-def}
\set{G}{j,k}&\triangleq \left\{(i_1,\ldots,i_m):\mathrm{coeff}\left(\int_0^{x_k} g_j(x_1,\ldots,x_{k-1},z,x_{k+1},\ldots x_m)dz,\prod_{l=1}^m x_l^{i_l}\right)\neq 0\right\}, \quad \text{for }j,k=1,\ldots,m
\end{align}
}
\end{figure*}

{\definition\label{def:Sets}
Let $\set{F}{j,k}$ and $\set{G}{j,k}$ be the sets defined by \eqref{eq:Fjk-def} and \eqref{eq:Gjk-def} respectively.
Then, the sets of non-zero coefficients of the functions $F(\yvec;\eps)$ and $G(\xvec)$, denoted by $\mathcal{S}^{F}$ and $\mathcal{S}^{G}$, respectively, are
\begin{align}
\label{eq:SetsFG}
\mathcal{S}^{F} \triangleq  \bigcup_{s=1}^m \bigcup_{(j,s) \in \set{D}{s}} \set{F}{j,s}, \quad\quad\quad \mathcal{S}^{G} \triangleq \bigcup_{s=1}^m \bigcup_{(j,s) \in \set{D}{s}} \set{G}{j ,s}
\end{align}
where
\begin{align}
\set{D}{s} \triangleq \{ (j,s): d_{js} \not = 0, \ 1 \le j \le m\}
\end{align}
for any $1 \le s \le m$.
}

\begin{remark} \label{remark-1}
For any $j=1, \ldots, m$ and $k=1, \ldots, m$, the sets $\set{F}{j,k}$ and $\set{G}{j,k}$ can be equivalently defined as
\begin{align*}
\set{F}{j,k} & \triangleq\left( \mathcal{S}_j^f + \boldsymbol{e}_k \right)\\
\set{G}{j,k} & \triangleq\left( \mathcal{S}_j^g + \boldsymbol{e}_k \right)
\end{align*}
where $\boldsymbol{e}_k$ is the standard basis vector of length $m$ with a one in the $k$th position and zero elsewhere, and the summation of a set $\mathcal S$ with a vector $\boldsymbol{e}_k$ is performed element-by-element.
\end{remark}

Given the expressions from Remark \ref{remark-1}, it is easy to verify that $|\set{F}{j,k}| \le |\set{f}{j}|+1$ and $|\set{G}{j,k}| \le |\set{g}{j}|+1$.


\begin{lemma}
\label{lemma:conf-conservative}
 For any coefficients $d_{ij}$, 
\begin{align}
\label{eq:coefficientsFG}
 \set{F}{} \subseteq \bigcup_{j,k=1}^m \set{F}{j,k},\quad\quad\quad 
 \set{G}{}\subseteq\bigcup_{j,k=1}^m \set{G}{j,k}.
\end{align}
Moreover, 
\begin{itemize}
\item $\set{F}{} = \bigcup_{j,k=1}^m \set{F}{j,k}$ and $\set{G}{}=\bigcup_{j,k=1}^m \set{G}{j,k}$ if $d_{ij} \not = 0$ for all $i,j$ (i.e., if $\Dvec$ does not have zero entries); 
\item $\set{F}{} = \bigcup_{j=1}^m \set{F}{j,j}$ and $\set{G}{}=\bigcup_{j=1}^m \set{G}{j,j}$ if $d_{ij} \not =0$ for $i=j$ and $0$ otherwise (i.e., if $\Dvec$ is diagonal).
\end{itemize}
\end{lemma}
\begin{IEEEproof}
The proof follows from direct calculation and is therefore omitted.
\end{IEEEproof}

We now address the existence of the potential function for the general case of a sparse-graph ensemble over the BEC defined by a vector recursion with multivariate polynomial functions $\fvec(\yvec;\varepsilon)$ and $\gvec(\xvec)$, without imposing any constraint on the matrix $\Dvec$. The special form of $\Dvec$ is discussed in Section~\ref{sec:ChoiceD}. We can state the following theorem.
{\theorem
\label{theorem:existence-F-G}
Consider a sparse-graph code ensemble, used for transmission over the BEC, whose density evolution updates $ \fvec(\yvec;\varepsilon)$ and $\gvec(\xvec)$ satisfy Lemma~\ref{lem:Lemma1}. Let the corresponding potential function be
\begin{align*}
U(\xvec; \varepsilon) 
\triangleq \gvec(\xvec)\Dvec \xvec\T -G(\xvec)-F(\gvec(\xvec);\varepsilon)
\end{align*}
with functions $F(\yvec; \varepsilon)$ and $G(\xvec)$ satisfying $F'(\yvec; \varepsilon) = \fvec(\yvec;\varepsilon) \Dvec$, 
$G'(\xvec) = \gvec(\xvec) \Dvec$, $F(\zerovec)=0$ and $G(\zerovec)=0$, for some $m\times m$ matrix $\Dvec$. Then, 
for any value of $\varepsilon$, $F(\yvec; \varepsilon)$ and $G(\xvec)$ exist (hence $U(\xvec; \varepsilon)$ exists) if there exist sets of values $\{d_{ j s} \}$, $\{{\varphi}_{(i_1,\ldots,i_m)}\}$ and $\{{\mu}_{(k_1,\ldots k_m)} \}$ that satisfy the following system of linear equations,
\begin{align}
\label{eq:system}
\begin{cases}
i_s {\varphi}_{(i_1,\ldots, i_s, \ldots,i_m)} = \sum_{j=1}^m d_{ j s} {\phi}^{(j)}_{(i_1,\ldots i_s-1, \ldots,i_m)}\\
k_t {\mu}_{(k_1,\ldots, k_t, \ldots,k_m)} = \sum_{j=1}^m d_{j t}  {\gamma}^{(j)}_{(k_1,\ldots,k_t-1,\ldots k_m)}
\end{cases}
\end{align}
for all possible $m$-tuples $(i_1,\ldots,i_m)$ and $(k_1,\ldots,k_m)$, and all values of $i_s$ and $k_t$ for $1\le s \le m$ and $1 \le t\le m$.
The coefficients $\phi$ and $\gamma$ in (\ref{eq:system}) are given by
\begin{align}
\label{eq:system-f-g}
{\phi}^{(j)}_{(i_1, \ldots,i_m)} &= \mathrm{coeff}(f_j(\yvec; \varepsilon),y_1^{i_ 1}\cdots y_m^{i_m}) \\
{\gamma}^{(j)}_{(k_1,,\ldots k_m)} &= \mathrm{coeff}(g_j(\xvec),x_1^{i_ 1}\cdots x_m^{i_m}).
\end{align}
}

\begin{IEEEproof}
Functions $F(\yvec;\varepsilon)$ and $G(\xvec)$ are multivariate polynomials, which can be written as
\begin{align}
\label{eq:FF}
F(\yvec;\varepsilon)&= \sum_{(i_1,\ldots,i_m)\in\set{F}{}} 
{\varphi}_{(i_1,\ldots,i_m)} y_1^{i_1}\ldots y_m^{i_m}\\
\label{eq:GG}
G(\xvec)&=\sum_{(k_1,\ldots,k_m)\in\set{G}{}}{\mu}_{(k_1,\ldots,k_m)} x_1^{k_1}\ldots x_m^{k_m}
\end{align}
where $\set{F}{}$ and $\set{G}{}$ are the coefficient sets given in Definition~\ref{def:Sets} and Lemma~\ref{lemma:conf-conservative}.

We can also write $f_j(\yvec, \varepsilon)$ and $g_j(\xvec)$ in polynomial form as
\begin{align}
\label{eq:fAppB}
f_j(\yvec, \varepsilon)&=\sum_{ (i_1,\ldots,i_m) \in\setf{j}}{\phi}^{(j)}_{(i_1,\ldots,i_m)} y_1^{i_1}\ldots y_m^{i_m}\\
\label{eq:gAppB}
g_j(\xvec)&=\sum_{(k_1,\ldots,k_m) \in \setg{j}}{\gamma}^{(j)}_{(k_1,\ldots,k_m)} x_1^{k_1}\ldots x_m^{k_m}
\end{align}
where $\setf{j}$ and $\setg{j}$ are the coefficient sets given in (\ref{eq:coefff})--(\ref{eq:coeffg}). Now, using (\ref{eq:FF})--(\ref{eq:GG}) and (\ref{eq:fAppB})--(\ref{eq:gAppB}) in (\ref{eq:F1})--(\ref{eq:G1}), we obtain the relationship between $\phi$'s, $\gamma$'s, $d$'s, $\varphi$'s and $\mu$'s, given by (\ref{eq:system}).
\end{IEEEproof}

It is important to note that, given the coefficients $\phi$ and $\gamma$ (which are fixed and known), (\ref{eq:system}) is a fixed homogenous linear system with respect to a single vector containing all the $d_{ij}$ coefficients, all the $\varphi$ coefficients, and all the $\mu$ coefficients.  We remark that (\ref{eq:system}) is a structured system of equations, so the existence of a solution does not follow from dimensional arguments: even though there might exist more equations than free variables $d_{ij}$, $\varphi$ and $\mu$, a solution might still exist as some of the equations in (\ref{eq:system}) are usually linearly dependent. 
Moreover, to show the existence of a solution for (\ref{eq:system}), one should rather consider the matching of the sets $\set{F}{}$ and $\set{G}{}$ with the sets $\set{F}{j,k}$ and $\set{G}{j,k}$. 
As this matching happens by means of the sets $\set{D}{s}$ (i.e., using nonzero coefficients in the matrix $\Dvec$), the existence of the solution depends on the form of $\Dvec$. The existence of a solution of (\ref{eq:system}) for various forms of $\Dvec$ is addressed in the next section.

Furthermore, note that, by definition, the coefficients $\phi$ depend on $\varepsilon$. Therefore, the coefficients $\varphi$ will also be functions of $\varepsilon$. We also remark that if we consider $\fvec(\yvec;\eps)$ and $\gvec(\xvec)$ defined by spatially-coupling a single-system DE, the discussion above also applies. The main difference is that the dimension of the system of equations will be larger.



\subsection{Necessary Condition on the Existence of $U(\Xvec; \varepsilon)$}
\label{sec:ChoiceD}

In Theorem~\ref{theorem:general-condition} below, we give a necessary condition for the existence of the potential function for any matrix $\Dvec$. We then consider the condition for a diagonal matrix $\Dvec$ and for a matrix $\Dvec$ with strictly positive entries. We will make use of the following definition. 

\begin{definition}
For a set $\mathcal S$ of vectors of length $m$ and a vector $\boldsymbol{e}_k$, let the subtraction operation ${\mathcal S} - \boldsymbol{e}_k$ be defined as
\begin{align*}
{\mathcal S} - \boldsymbol{e}_k =& \{ (i_1, \ldots, i_k, \ldots, i_m): \\ &(i_1, \ldots, i_k-1, \ldots, i_m) \in {\mathcal S}
 \text{ and } i_k>0  \}.
\end{align*}
\end{definition}

\begin{theorem} \label{theorem:general-condition} 
If the system of equations (\ref{eq:system}) exists then, for all $s=1,\ldots,m$ and $j = 1, \ldots, m$, it holds that
\begin{align} \label{eq:conditionsaa1}
  \left( \bigcup_{(i,s)\in \set{D}{s}} \set{F}{i,s} - \boldsymbol{e}_j \right) \subseteq  \left(\bigcup_{(j,s')\in \set{D}{}} \set{f}{j} \right)
\end{align}
and
\begin{align} \label{eq:conditionsaa2}
\left( \bigcup_{(i,s)\in \set{D}{s}} \set{G}{i,s}- \boldsymbol{e}_j \right) \subseteq  \left(\bigcup_{(j,s')\in \set{D}{}} \set{g}{j} \right)
\end{align}
where $\set{D}{} = \bigcup_{k=1}^m \set{D}{k}$.
Equivalently, \eqref{eq:conditionsaa1}--\eqref{eq:conditionsaa2} can be written as
\begin{align}
  \left( \bigcup_{(i,s)\in \set{D}{s}} \left(\set{f}{i} +  \boldsymbol{e}_s \right) - \boldsymbol{e}_j \right) \subseteq  \left(\bigcup_{(j,s')\in \set{D}{}} \set{f}{j} \right) 
\end{align}
and  
\begin{align}
\left( \bigcup_{(i,s)\in \set{D}{s}} \left( \set{g}{i} +  \boldsymbol{e}_s \right)- \boldsymbol{e}_j \right) \subseteq  \left(\bigcup_{(j,s')\in \set{D}{}} \set{g}{j} \right).
\end{align}
\end{theorem}
\begin{IEEEproof} Consider the sets $\set{G}{i,s}$.
Assume that there exist an element $\avec= (i_1,\ldots,i_s+1,\ldots,i_j-1,\ldots,i_m)$ such that $\avec \in\left( \bigcup_{(i,s)\in \set{D}{s}} \left( \set{g}{i} +  \boldsymbol{e}_s \right)- \boldsymbol{e}_j \right)$ but $\avec \not\in \left(\bigcup_{(j,s')\in \set{D}{}} \set{g}{j} \right)$. Since $\avec\in\left( \bigcup_{(i,s)\in \set{D}{s}} \left( \set{g}{i} +  \boldsymbol{e}_s \right)- \boldsymbol{e}_j \right)$, then $\tilde{\avec} =\avec+\evec_j\in  \set{G}{}$. We get $\tilde{\avec} = (i_1,\ldots ,i_s+1,\ldots,i_j, \ldots,i_m)$, i.e., the monomial $x_1^{i_1}\cdots x_{i}^{i_s+1}\cdots x_j^{i_j}\cdots x_m^{i_m}$ appears in $G(\xvec)$. But now $\derivF{G(\xvec)}{x_j}$ contains the monomial $x_1^{i_1}\cdots x_{i}^{i_s+1}\cdots x_j^{i_j-1}\cdots x_m^{i_m}$, which is a contradiction, since by assumption $$(i_1,\ldots,i_s+1,\ldots,i_j-1,\ldots,i_m)\not\in  \left(\cup_{(j,s')\in \set{D}{}}\set{g}{j} \right).$$ The same reasoning holds for the sets $\set{F}{i,s}$.
\end{IEEEproof}
\begin{corollary}
\label{cor:CondDiagonal}
For a diagonal $\Dvec$, the condition is
\begin{align}
\label{eq:CondDiagonal}
 \left(\set{f}{i} +  \boldsymbol{e}_i \right) - \boldsymbol{e}_j \in \set{f}{j} \quad \text{and} \quad   \left( \set{g}{i} +  \boldsymbol{e}_s \right)- \boldsymbol{e}_j  \in \set{g}{j}
\end{align}
for $1\le i \le m, \ 1 \le j \le m$. 
\end{corollary}

\begin{corollary}
For a strictly positive $\Dvec$, the conditions are
\begin{align} \label{eq:conditions-positive-D}
 & \left( \bigcup_{s=1}^m \left(\set{f}{i} +  \boldsymbol{e}_s \right) - \boldsymbol{e}_j \right) \in \cup_{j=1}^m \set{f}{j} \\
 & \left(  \bigcup_{s=1}^m \left( \set{g}{i} +  \boldsymbol{e}_s \right)- \boldsymbol{e}_j \right) \in \cup_{j=1}^m \set{g}{j}
\end{align}
for $1\le i \le m, \ 1 \le j \le m$. 
\end{corollary}

In the following, we give an example where the necessary condition for a diagonal matrix $\Dvec$ is satisfied and we compute the potential function by solving the system of linear equations \eqref{eq:system}.
\begin{example}[Bilayer LDPC code for the relay channel \cite{Raz07}] We compute the potential function for a regular bilayer LDPC code with parameters $(\ell_1,\ell_2,r_1,r_2)$ for transmission over the binary erasure relay channel with equal erasure probabilities $\varepsilon$. $\ell_1$ and $\ell_2$ are the variable node degrees for the first and second layer, respectively. $r_1$ and $r_2$ are the check node degrees for the first and second layer. The extension to irregular codes is straightforward. The DE for the bilayer code is given by
\begin{align}
\label{eq:DEex}
\begin{cases}
x_1=\varepsilon y_1^{\ell_1-1}y_2^{\ell_2},~~
y_1=1-(1-x_1)^{r_1-1},&\text{(first layer)}\\
x_2=\varepsilon y_1^{\ell_1}y_2^{\ell_2-1},~~
y_2=1-(1-x_2)^{r_2-1},&\text{(second layer)}
\end{cases}
\end{align}
where $x_1$ is the erasure probability of messages from variable nodes to check nodes in the first layer, and $y_1$ is the erasure probability of messages from check nodes in the first layer to variable nodes. Likewise, $x_2$ is the erasure probability of messages from variable nodes to check nodes in the second layer, and $y_2$ is the erasure probability of messages from check nodes in the second layer to variable nodes. 

We write the corresponding vector functions
\begin{align}
\label{eq:VFex}
\fvecy &= (f_1,f_2)=(\varepsilon y_1^{\ell_1-1}y_2^{\ell_2},\varepsilon y_1^{\ell_1}y_2^{\ell_2-1})\\
\label{eq:VGex}
\gvecx &=(g_1,g_2)=(1-(1-x_1)^{r_1-1},1-(1-x_2)^{r_2-1})
\end{align}
and obtain the following coefficient sets,
\begin{align}
\set{f}{1}&=\{(\ell_1-1,\ell_2)\},~~~~\set{g}{1}=\{(0,0),(1,0),\ldots,(r_1-1,0)\},\nonumber\\
\set{f}{2}&=\{(\ell_1,\ell_2-1)\},~~~~\set{g}{2}=\{(0,0),(0,1),\ldots,(0,r_2-1)\}.\nonumber
\end{align}
It is easy to verify that the necessary condition in Corollary~\ref{cor:CondDiagonal} holds.

Assume a diagonal matrix 
$\Dvec = \left[\begin{array}{cc}d_{11} & 0 \\0 & d_{22}\end{array}\right].$
We obtain
\begin{align*}
\set{F}{}&\triangleq\bigcup_{j=1}^m \set{F}{j,j}=\bigcup_{j=1}^2 \left(\set{f}{j}+ \boldsymbol{e}_j\right)=\{(\ell_1,\ell_2)\}\\
 \set{G}{}&\triangleq\bigcup_{j=1}^m \set{G}{j,j}=\bigcup_{j=1}^2 \left(\set{g}{j}+ \boldsymbol{e}_j\right)\\ &=\{(1,0),\ldots,(r_1,0),(0,1),\ldots,(0,r_2)\}.
\end{align*}

To derive $\Fvecy$, $\Gvecx$, and $\Dvec$, we first write the system of linear equations, as explained above, as
\begin{align}
\label{eq:systemex}
\begin{cases}
\ell_1\varphi_{(\ell_1,\ell_2)}&=d_{11}\phi^{(1)}_{(\ell_1-1,\ell_2)}\\
\ell_2\varphi_{(\ell_1,\ell_2)}&=d_{22}\phi^{(2)}_{(\ell_1,\ell_2-1)}\\
i\mu_{(i,0)}&=d_{11}\gamma^{(1)}_{(i-1,0)}\\
i\mu_{(0,i)}&=d_{22}\gamma^{(2)}_{(0,i-1)}
\end{cases}
\end{align}
for $i=1,\ldots,r_1$.

From (\ref{eq:VFex})-(\ref{eq:VGex}), and using the binomial formula to expand $(1-x)^a$, we obtain the coefficients
\begin{align*}
\phi^{(1)}_{(\ell_1-1,\ell_2)}=\varepsilon, &\quad \quad \phi^{(2)}_{(\ell_1,\ell_2-1)}=\varepsilon\\
\gamma^{(1)}_{(0,0)}=0, &\quad \quad \gamma^{(1)}_{(i,0)}= -\binom{r_1-1}{i}(-1)^i \\
 \gamma^{(2)}_{(0,1)}=0, &\quad \quad \gamma^{(2)}_{(0,i)}= -\binom{r_2-1}{i}(-1)^i.
\end{align*}
Using these coefficients in (\ref{eq:systemex}) and solving the system of linear equations, we obtain\footnote{The example presents one solution out of many existing ones. In fact, for any $a>0$ and $d_{11}=a \ell_1$, $d_{22}=a \ell_2$ and $\varphi_{(\ell_1,\ell_2)}=\varepsilon/a$ induce a solution which satisfies \eqref{eq:systemex}.} $d_{11}=\ell_1$, $d_{22}=\ell_2$, $\varphi_{(\ell_1,\ell_2)}=\varepsilon$, $\mu_{(1,0)}=\mu_{(0,1)}=0$, and, for $2 \le i <r$,
\begin{align*}
\mu_{(i,0)}&=-\frac{d_{11}}{i}\binom{r_1-1}{i-1}(-1)^{i-1}\\
 \mu_{(0,i)}&=-\frac{d_{22}}{i}\binom{r_1-1}{i-1}(-1)^{i-1}.
\end{align*}

Finally, the functions $F(\yvec;\varepsilon)$ and $G(\xvec)$ are found as (see (\ref{eq:FF}) and (\ref{eq:GG}))
\begin{align*}
F(\yvec;\varepsilon)= \sum_{(i_1,i_2)\in\set{F}{}} {\varphi}_{(i_1,i_2)} y_1^{i_1} y_2^{i_2}=
\varphi_{(\ell_1,\ell_2)}y_1^{\ell_1}y_2^{\ell_2}=\varepsilon y_1^{\ell_1}y_2^{\ell_2}
\end{align*}
and
\begin{align}
\label{eq:Gexample}
G(\xvec)&=\sum_{(k_1,k_2)\in\set{G}{}}{\mu}_{(k_1,k_2)} x_1^{k_1} x_2^{k_2} \nonumber \\&=\sum_{i=1}^{r_1}\mu_{(i,0)}x_1^i+\sum_{i=1}^{r_2}\mu_{(0,i)}x_2^i.
\end{align}

The first summation term in right hand side of (\ref{eq:Gexample}) becomes
\begin{align*}
\sum_{i=1}^{r_1}\mu_{(i,0)}x_1^i &= \mu_{(1,0)} -\sum_{i=2}^{r_1} \frac{d_{11}}{i}\binom{r_1-1}{i-1}(-1)^{i-1}x_1^i\\
&=\frac{d_{11}}{r_1}\sum_{i=2}^{r_1} \binom{r_1}{i}(-1)^{i}x_1^i\\
&=\frac{d_{11}}{r_1}\left(-1+r_1x_1+\sum_{i=0}^{r_1} \binom{r_1}{i}(-1)^{i}x_1^i\right)\\
&\stackrel{(a)}{=}d_{11}\left(x_1+\frac{(1-x_1)^{r_1}-1}{r_1}\right)
\end{align*}
where in $(a)$ we used the binomial formula.

Developing the second term in the summation (\ref{eq:Gexample}) in a similar way, and setting $d_{11}=\ell_1$ and $d_{22}=\ell_2$, we finally obtain
\begin{align}
\label{eq:Gexample2}
\Gvecx &=\ell_1\left(x_1+\frac{(1-x_1)^{r_1}-1}{r_1}\right) \nonumber \\
&+\ell_2\left(x_2+\frac{(1-x_2)^{r_2}-1}{r_2}\right).
\end{align}
It is easy to verify that $\Fvecy$ and $\Gvecx$ satisfy $F(\zerovec;\varepsilon)=0$,  $G(\zerovec)=0$, $F'(\yvec; \varepsilon) = \fvec(\yvec;\varepsilon) \Dvec$, and
$G'(\xvec) = \gvec(\xvec) \Dvec$.\demo
\end{example}

%

\subsection{Existence of $U(\xvec;\eps)$ for $(\dl,\dr,m)$ and $(\dl,\dr,m, L, \dw)$ Ensembles}
\label{sec:F-and-G-nb}

In \cite{pfister-itw,pfister-vector-itw}, the potential function was defined using a diagonal matrix $\Dvec$. It is easy to verify that all the examples in \cite{pfister-itw} (noisy Slepian-Wolf problem with erasures, LDPC codes over the erasure multiple access channel, and protograph codes over the BEC) satisfy the necessary condition in Corollary~\ref{cor:CondDiagonal}. However, in the case of nonbinary LDPC codes, the following proposition is true.
\begin{proposition} 
\label{prop:prop} 
For the $(\dl,\dr,m)$ nonbinary LDPC code ensemble, if $\Dvec$ is a diagonal matrix, the solution to \eqref{eq:system} does not exist.
\end{proposition}
\begin{IEEEproof}
First consider the variable node operation in (\ref{eq:bij1}) using (\ref{eq:boxdot}). Consider $\set{f}{m}$, the set of nonzero coefficients of $f_m$ (from \eqref{eq:Map2}, $f_m=f_{\circ m}$). From (\ref{eq:boxdot}), $[\avecc\boxdot\bvecc]_m=V^m_{m,m,m}a_{\circ m}b_{\circ m}$. Therefore, $f_m=p_m(V^m_{m,m,m})^{\dl-1}y_m^{\dl-1}$, and $\set{f}{m}=\{(0,\ldots,0,\dl-1)\}$. Developing $[\avecc\boxdot\bvecc]_{m-1}$ in (\ref{eq:bij1}) and using $[\avecc\boxdot\bvecc]_m$ earlier, it is also easy to see that 
\begin{multline*}
\set{f}{m-1}=\{(0,\ldots,0,\dl-1,0),\\(0,\ldots,0,\dl-2,1), \ldots,(0,\ldots,0,\dl-1)\}.
\end{multline*} 
Therefore, the condition \eqref{eq:CondDiagonal} is not verified. A similar proof holds for the check node operation.
\end{IEEEproof}

As a result of Proposition~\ref{prop:prop}, for nonbinary LDPC codes, the potential function does not exist for a diagonal matrix $\Dvec$. Thus, one should consider a more general form of $\Dvec$. In the following, we show that, indeed, the potential function exists if a more general form of $\Dvec$ is considered. We first give an example of the existence and computation of the potential function for the $(2,3,3)$ nonbinary LDPC code, and then discuss the existence of a solution to the system of linear equations \eqref{eq:system} for nonbinary LDPC codes in general.

\begin{example}[Existence and calculation of the potential function for the $(2,3,3)$ nonbinary LDPC code]
\label{ex:ExNB}
We have $\xvecO=(x_{\circ 0},x_{\circ 1},x_{\circ 2})$, $\yvecO=(y_{\circ 0},y_{\circ 1},y_{\circ 2})$, $\pvecO=((1-\varepsilon)^2,2\varepsilon(1-\varepsilon),\varepsilon^2)$, $\fvecO(\yvecO;\varepsilon)=\pvecO(\varepsilon)\boxdot\yvecO$ and $\gvecO(\xvecO)=\xvecO\boxtimes\xvecO$. Moreover, $\xvec=(x_1,x_2)$, $\yvec=(y_1,y_2)$ and $\fvec(\yvec;\varepsilon)=(f_1,f_2)$. Using (\ref{eq:boxdot}) we obtain
\begin{align*}
f_{\circ 0} &= 1-f_{\circ 1}-f_{\circ 2}\\
f_{\circ 1} &= V^2_{1,1,1}p_{\circ 1}y_{\circ 1}+V^2_{1,2,1}p_{\circ 1}y_{\circ 2}+V^2_{2,1,1}p_{\circ 2}y_{\circ 1} \\
&=\left(\frac{2}{3}\varepsilon(1-\varepsilon)+\varepsilon^2\right)y_{\circ 1}+2\varepsilon(1-\varepsilon)y_{\circ 2}\\
f_{\circ 2} &= V^2_{2,2,2}p_{\circ 2}y_{\circ 2}=\varepsilon^2y_{\circ 2}.
\end{align*}
Using \eqref{eq:Map1} and \eqref{eq:Map2} we obtain,
\begin{align*}
\fvec=(f_1,f_2)=\left(\frac{2\eps+\eps^2}{3}y_1+\frac{4\eps-4\eps^2}{3}y_2,\eps^2y_2\right).
\end{align*}
It follows that $S_1^f=\{(1,0),(0,1)\}$ and $S_2^f=\{(0,1)\}$. 

\begin{figure*}
{\footnotesize
\begin{align}
\label{eq:fvec-long}
\fvec(\yvec;\eps)\Dvec=& \left(\frac{(2\eps+\eps^2)d_{11}}{3}y_1+\frac{(4\eps-4\eps^2)d_{11}+3\eps^2d_{21}}{3}y_2,\frac{(2\eps+\eps^2)d_{12}}{3}y_1+\frac{(4\eps-4\eps^2)d_{12}+3\eps^2d_{22}}{3}y_2\right)
\\
\label{eq:gvec-long}
\gvec(\xvec)\Dvec=& \left(2d_{11}x_1+\frac{2d_{21}-3d_{11}}{3}x_1^2-\frac{4d_{21}}{3}x_1x_2+2d_{21}x_2-\frac{d_{21}}{3}x_2^2,
 2d_{12}x_1+\frac{2d_{22}-3d_{12}}{3}x_1^2-\frac{4d_{22}}{3}x_1x_2+2d_{22}x_2-\frac{d_{22}}{3}x_2^2 \right)
\end{align}
}
\end{figure*}

Similarly, we obtain
\begin{align*}
g_{\circ 0} &= 1-g_{\circ 1}-g_{\circ 2}\\
g_{\circ 1} &
=2x_{\circ 1}-\frac{5}{3}x_{\circ 1}^2-2x_{\circ 1}x_{\circ 2}\\
g_{\circ 2} &= \frac{2}{3}x_{\circ 1}^2-\frac{4}{3}x_{\circ 1}x_{\circ 2}+2x_{\circ 2}-\frac{1}{3}x_{\circ 2}^2
\end{align*}
and\begin{align*}
\gvec=(g_1,g_2)=\left(2x_1-x_1^2,\frac{2}{3}x_1^2-\frac{4}{3}x_1x_2+2x_2-\frac{1}{3}x_2^2\right).
\end{align*}
It follows that $S_1^g=\{(1,0),(2,0)\}$ and $S_2^g=\{(2,0),(1,1),(0,1),(0,2)\}$.

With the sets above,  \eqref{eq:CondDiagonal} is not satisfied. Therefore, the potential function does not exist for a diagonal matrix $\Dvec$. However, the potential function exists if a matrix with non-zero elements is used. In this case the sets $\mathcal{S}^F$ and $\mathcal{S}^G$ are (see (\ref{eq:SetsFG})),
\begin{align*}
\mathcal{S}^F &= \{(2,0),(1,1),(0,2)\}\\
\mathcal{S}^G &= \{(2,0),(3,0),(1,1),(1,2),(2,1),(0,2),(0,3)\}.
\end{align*}
Using these sets, we can write the functions $F(\yvec;\eps)$ and $G(\xvec)$ as (cf.~\eqref{eq:FF} and \eqref{eq:GG})
\begin{align*}
F(\yvec;\eps) &= \varphi_{(2,0)}y_1^2+\varphi_{(1,1)}y_1y_2+\varphi_{(0,2)}y_2^2\\
G(\xvec) &= \mu_{(2,0)}x_1^2+\mu_{(3,0)}x_1^3+\mu_{(1,1)}x_1x_2+\mu_{(1,2)}x_1x_2^2 \\ &~~~+\mu_{(2,1)}x_1^2x_2+\mu_{(0,2)}x_2^2+\mu_{(0,3)}x_2^3.
\end{align*}
We also obtain
\begin{align*}
F'(\yvec;\eps)&=\derivF{F(\yvec;\eps)}{\yvec} \\&= \left(2\varphi_{(2,0)}y_1+\varphi_{(1,1)}y_2,\varphi_{(1,1)}y_1+2\varphi_{(0,2)}y_2\right)\\
G'(\xvec)&=\derivF{G(\xvec)}{\xvec}\\&=(2\mu_{(2,0)}x_1+3\mu_{(3,0)}x_1^2+\mu_{(1,1)}x_2+\mu_{(1,2)}x_2^2 \\ &~~~~+2\mu_{(2,1)}x_1x_2,
\mu_{(1,1)}x_1+2\mu_{(1,2)}x_1x_2 \\ & ~~~~+\mu_{(2,1)}x_1^2+2\mu_{(0,2)}x_2+3\mu_{(0,3)}x_2^2).
\end{align*}

On the other hand, $\fvec(\yvec;\eps)\Dvec$ and $\gvec(\xvec)\Dvec$ are given by \eqref{eq:fvec-long} and \eqref{eq:gvec-long}, respectively, at the top of the page.

We can now write the system of linear equations (\ref{eq:system}) as
\begin{align*}
2\varphi_{(2,0)}=&\frac{(2\eps+\eps^2)d_{11}}{3},  &\varphi_{(1,1)} =&\frac{(4\eps-4\eps^2)d_{11}+3\eps^2d_{21}}{3},\nonumber\\
\varphi_{(1,1)}=&\frac{(2\eps+\eps^2)d_{12}}{3},   & 2\varphi_{(0,2)} =&\frac{(4\eps-4\eps^2)d_{12}+3\eps^2d_{22}}{3},\nonumber\\
2\mu_{(2,0)}=&~~2d_{11},  & 3\mu_{(3,0)} =&\frac{2d_{21}-3d_{11}}{3},\nonumber\\
\mu_{(1,1)}=&~~2d_{21},  & \mu_{(1,2)}=&-\frac{d_{21}}{3},\nonumber\\
2\mu_{(2,1)}=&-\frac{4d_{21}}{3},   &\mu_{(1,1)}=&~~2d_{12},\nonumber\\ 
2\mu_{(1,2)} =& -\frac{4d_{22}}{3},  &\mu_{(2,1)}=&\frac{2d_{22}-3d_{12}}{3},\nonumber\\
2\mu_{(0,2)}=& 2d_{22},  & 3\mu_{(0,3)}=&-\frac{d_{22}}{3}.
\end{align*}
Note that solving the system above is equivalent to solve the following subsystem of linear equations,
\begin{align*}
\begin{cases}
(2 \varepsilon+\varepsilon^2) d_{12} &= (4\varepsilon-4\varepsilon^2)d_{11} + 3 \varepsilon^2 d_{21}\\
2 d_{21}&= 2 d_{12}\\
-\frac{4}{6}d_{22} &= -\frac{1}{3}d_{21}
\end{cases}
\end{align*}
This subsystem gives us the solution
\begin{align*}
d_{12} &= 2d_{11},\quad & d_{21} &=2 d_{11},\quad  & d_{22} &= d_{11}
\end{align*}
and the induced coefficients $\varphi$ and $\mu$ are then obtained as
\begin{align*}
\varphi_{(2,0)} =&\frac{2\eps+\eps^2}{6}d_{11}, &
\mu_{2,0} =& ~d_{11},  & 
\mu_{1,2} =&  -\frac{2}{3}d_{11},\\ 
\varphi_{(1,1)} =& \frac{4\eps+2\eps^2}{3}d_{11}, & \mu_{3,0} =&~  \frac{1}{9}d_{11},  &\mu_{1,1} =& ~4d_{11}\\ 
\varphi_{(0,2)} =& \frac{8\eps-5\eps^2}{6}d_{11},&
\mu_{2,1} =&  -\frac{4}{3}d_{11},\quad &\mu_{0,2} =&~  d_{11},\\ 
&&\mu_{0,3} =&  -\frac{1}{9}d_{11}.
\end{align*}
Therefore, the system has an infinite number of solutions. Choosing the free parameter $d_{11}=1$ we finally obtain
\begin{align*}
F(\yvec;\eps) &= \frac{2\eps+\eps^2}{6}y_1^2+\frac{4\eps+2\eps^2}{3}y_1y_2+\frac{8\eps-5\eps^2}{6}y_2^2\\
G(\xvec) &= x_1^2+\frac{1}{9}x_1^3+4x_1x_2-\frac{2}{3}x_1x_2^2-\frac{4}{3}x_1^2x_2+x_2^2-\frac{1}{9}x_2^3.
\end{align*}
It is easy to verify that this solution satisfies $F'(\yvec;\eps)=\fvec(\yvec;\eps)\Dvec$ and $G'(\xvec)=\gvec(\xvec)\Dvec$. Note also that the resulting $\Dvec$ is symmetric. \demo
\end{example}

}

We now discuss the general case of nonbinary LDPC codes. Assume that the necessary condition on the existence of functions $F(\yvec;\eps)$ and $G(\xvec)$ is satisfied. We would like to determine whether \eqref{eq:system} has always a solution. To do so, we analyze the form of the system of equations \eqref{eq:system} for nonbinary LDPC codes.
\begin{lemma}
For nonbinary $(\dl, \dr, m)$ LDPC codes,
\begin{align} \label{eq:|setFG|-1}
|\set{F}{}| &= \sum_{n=3}^{\dl} \sum_{t=1}^{n} { {n-1} \choose {t-1}} { m \choose t} \\
\label{eq:|setFG|-2}
|\set{G}{}| &= \sum_{n=2}^{\dr} \sum_{t=1}^{n} {{n-1} \choose  {t-1}}{m \choose t}.
\end{align}
Moreover, the number of equations in \eqref{eq:system}, containing ${\varphi}_{(i_1,\ldots,i_m)}$ and ${\mu}_{(i_1,\ldots,i_m)}$ is respectively
\begin{align}
\label{eq:Nfg-1}
N_{\varphi} &=\sum_{n=3}^{\dl} \sum_{t=1}^{n} t {{n-1} \choose{t-1}} {m \choose t}\\
\label{eq:Nfg-2}
N_{\mu} &=\sum_{n=2}^{\dr} \sum_{t=1}^{n} t {{n-1}\choose {t-1}} { m \choose t}.
\end{align}
\end{lemma}
\begin{IEEEproof}
Consider \eqref{eq:FF}. It defines a multivariate polynomial $F(\yvec;\eps)$ which, by construction, contains all monomials $y_1^{i_1}\cdots y_m^{i_m}$ such that 
$$3 \le \sum_{s=1}^m i_s \le \dl.$$
Therefore, in order to find the number of elements in $\set{F}{}$, we count the number of vectors $(i_1,\ldots,i_m)\in \set{F}{}$ of length $m$ with $t$ nonzero entries, and $\sum_{s=1}^m i_s=n$, for $3 \le n \le \dl$ and $1 \le t \le m$,. As for some fixed value of $n$, the number of vectors with exactly $t$ nonzero entries and $\sum_{s=1}^m i_s=n$ is given by the binomial coefficient $ {n-1} \choose {t-1}$, and these $t$ nonzero entries can be placed in a vector of length $m$ in $ {m} \choose {t}$ various ways, the expression for $|\set{F}{}|$ follows directly. 
The same reasoning holds for $|\set{G}{}|$, with the only difference that $G(\xvec)$ is constructed in a way that it contains all monomials $x_1^{i_1}\cdots x_m^{i_m}$ such that $$2 \le \sum_{s=1}^m i_s \le \dr.$$
Now, to obtain \eqref{eq:Nfg-1}, it is sufficient to notice from \eqref{eq:system} that any variable ${\varphi}_{(i_1,\ldots,i_m)}$ having $t$ nonzero entries within its index vector $(i_1,\ldots, i_m)$ participates in exactly $t$ equations. The same holds for variables ${\gamma}_{(i_1,\ldots,i_m)}$.
\end{IEEEproof}

The system of linear equations \eqref{eq:system} has $|\set{F}{}| + |\set{G}{}| + \frac{m(m-1)}{2}$ free variables (which are the coefficients $\varphi$, $\mu$ and $d_{ij}$) and $N_{\varphi}+N_{\gamma}$ equations.
Therefore, if $\dl$, $\dr$ and $m$ are such that 
\begin{align}
\label{eq:cond-suff-simple}
|\set{F}{}| + |\set{G}{}| + \frac{m(m-1)}{2} \ge N_{\varphi}+N_{\gamma}
\end{align}
then \eqref{eq:system} would have a solution. 
Unfortunately, for almost all interesting choices of $\dl$, $\dr$ and $m$ this condition is not verified. 
However, the solution of \eqref{eq:system} does exist in most cases. This has been verified numerically, and is discussed in the following. The reason that a solution exists lies in the fact that, due to the structural properties of $V_{i,j,k}$ and $C_{i,j,k}$ in $\fvec(\yvec;\eps)$ and $\gvec(\xvec)$, the system \eqref{eq:system} contains many linearly dependent equations, so the condition \eqref{eq:cond-suff-simple} is too weak. To verify the existence of $F(\yvec;\eps)$ and $G(\xvec)$ for nonbinary LDPC codes, we verified numerically the existence of a solution for regular codes with degrees $(2,3)$, $(3,4)$, $(3,5)$ and $(3,6)$, for values of $m$ starting from $2$ until the largest value that was feasible to simulate (e.g., for $(3,4)$ LDPC codes we considered $m$ up to $20$).
For all the considered examples, the number of equations in \eqref{eq:system} was larger than the number of free variables. 
However, in all cases, similarly to Example~\ref{ex:ExNB}, after eliminating linearly dependent equations, we obtained a subsystem of $m^2-1$ equations with $m^2$ variables $d_{ij}$, which results in a symmetric matrix $\Dvec$, parametrized by $d_{11}$. Then, all coefficients $\varphi$ and $\mu$ are obtained as a function of $d_{11}$, as in Example~\ref{ex:ExNB}. The subsystems of linear equations for the $(3,4,2)$ and $(3,4,3)$ nonbinary LDPC codes are given in Table~\ref{table:examples}.
Our numerical results strongly suggest that, for $(\dl,\dr,m)$ nonbinary LDPC code ensembles, the functions $F(\yvec;\eps)$ and $G(\xvec)$ always exist, and the corresponding matrix $\Dvec$ is always symmetric. Choosing $d_{11}$ to be positive, the resulting matrix is positive, symmetric and invertible, properties that are used in Lemma~\ref{Lem:Potential} and subsequently in the proof of threshold saturation. 

\begin{table*}[!t]
\addtolength{\tabcolsep}{-0.0mm}
\caption{Subsystems of linearly independent equations from \eqref{eq:system}, for some values of $\dl$, $\dr$ and $m$.}
\vspace{-2ex}
\begin{center}\begin{tabular}{ccccc}
\toprule
$(\dl, \dr,m)$ & $|\set{F}{}| + |\set{G}{}|$ & $N_{\varphi}+N_{\gamma}$ & Equivalent subsystem of equations & Resulting $\Dvec$\\
\otoprule
 (3,4,2)& 16 & 24 & $\begin{cases}-\frac{4}{10}d_{21}+\frac{9}{10} d_{12}=d_{22}\\ -d_{21}+\frac{3}{2}d_{12}=d_{22} \\ 2 \varepsilon(1-\varepsilon)d_{11}=\varepsilon^2 d_{12} \end{cases}$ & $\left[\begin{array}{cc} d_{11}& 2d_{11}\\ 2 d_{11}&d_{11}\end{array}\right]$\\[0.5mm]
  (3,4,3)&41&75& $\begin{cases} 
 3 d_{21}=3d_{12}\\
 d_{31}=4_{d33}\\
 3d_{31}=3 d_{13}\\
 6 d_{32}=12 d_{33}\\
 d_{32}=d_{23}\\
 d_{21}+15 d_{33}=6 d_{22}\\
4 \varepsilon d_{11}=2\varepsilon d_{12}\\
8 \varepsilon d_{11}=2\varepsilon d_{13}
  \end{cases}$ & $\left[\begin{array}{ccc} d_{11}& 3 d_{11}&4 d_{11}\\3 d_{11}& 3d_{11}&2 d_{11} \\ 4d_{11}&2 d_{11}&d_{11} \end{array}\right]$\\[0.5mm]  
\bottomrule
\end{tabular} \end{center}
\label{table:examples} 
\vspace{-2ex}
\end{table*}


%

\section{Conclusion}
\label{sec:Conclusions}


We studied threshold saturation for nonbinary SC-LDPC codes when transmission takes place over the BEC. 
We used the proof technique based on the potential function $U(\xvec;\epsilon)$ for vector recursions, recently proposed by Yedla \textit{et al.} \cite{pfister-itw} and showed that threshold saturation occurs for nonbinary SCLDPC codes, under the condition of the existence of the potential function. Our proof is a non-straightforward extension of the proof in \cite{pfister-itw} to accommodate nonbinary SC-LDPC codes. In particular, during the proof of the threshold saturation, we have shown the following important facts:
\begin{itemize}
\item {\it Existence of a fixed point in the DE of nonbinary LDPC codes:}
In their probability vector form, the variable and check node updates $\fvec(\yvec;\eps)$ and $\gvec(\xvec)$ are not monotone with respect to the input variables. However, monotonicity can be shown by using an equivalent representation based on CCDF vectors. The property of monotonicity implies the existence of a fixed point in the DE equation of nonbinary LDPC ensembles, and also allows to use the proof technique of \cite{pfister-itw}.
\item {\it Existence and calculation of the potential function:} 
We derived a necessary condition on the existence of the potential function. Furthermore, we showed that, if it exists, the potential function can be obtained by finding the functions $F(\yvec;\eps)$ and $G(\xvec)$ as the solution of a system of linear equations.
\item {\it Use of a diagonal matrix $\Dvec$ in the definition of $U(\xvec;\epsilon)$:} In \cite{pfister-itw} $\Dvec$ is assumed to be diagonal. For many vector coupled systems a diagonal matrix $\Dvec$ satisfies the necessary condition for the existence of $U(\xvec;\epsilon)$. However, the condition is not verified in the case of nonbinary codes.
\item {\it Positive, symmetric and invertible form of $\Dvec$ and threshold saturation for nonbinary LDPC codes:} We showed numerically that, for multiple families of nonbinary LDPC code ensembles, $F(\yvec;\eps)$ and $G(\xvec)$ exist for
a positive, symmetric and invertible matrix $\Dvec$. In such cases,
we are able to prove threshold saturation following the same lines as the proof by Yedla \textit{et al.}.
Unfortunately, the general problem of the existence of a solution to the system of linear equations (and thus of the potential function) for arbitrary nonbinary LDPC code ensembles still remains open. However, based on our observations, we conjecture that threshold saturation occurs in general.
\end{itemize} 

\section*{Acknowledgemets}

The authors wish to thank Prof. Henry D. Pfister for suggesting the use of CCDF vectors to prove the monotonicity of the variable node and check node update functions. The authors would also like to thank Prof. Henry D. Pfister and the anonymous reviewers for the careful review of our paper.


\appendices

\section{Some Useful Results}
\label{app:UsefulResults}

Consider the Gaussian binomial coefficient in (\ref{e:gauss_bin_coeff}) and the definition of the coefficients$V_{ijk}$,
\begin{equation} \label{eq:Vijk}
V^m_{i,j,k}=\begin{cases} \frac{{\GaussBin}_{i,k} {\GaussBin}_{m-i,j-k} 
2^{(i-k)(j-k)}}{{\GaussBin}_{m,j}}, & \text{ if (C1) holds}\\
0, & \text{otherwise}
\end{cases}
\end{equation}
where the condition (C1) is 
\begin{align*}
\text{(C1) :} \quad & 0 \le i,j \le m \quad \text{ and } \\& \max(i+j-m,0) \le k \le \min(i,j).
\end{align*}

We give without proof the following lemmas.

{\lemma
Assume $m>0$. Then for all $0\le i\le m$,
\begin{align}
\frac{ G_{m, i+1}}{G_{m, i}} &= \frac{2^{m-i}-1}{2^{i+1}-1} \label{eq:1} \\
\frac{ G_{m+1, i}}{G_{m, i}} &= \frac{2^{m+1}-1}{2^{m-i+1}-1} \label{eq:2} \\
\frac{ G_{m+1, i+1}}{G_{m, i}} &= \frac{2^{m+1}-1}{2^{i+1}-1}. \label{eq:3}
\end{align}
}

{\lemma \label{lemma:3} Let $i,j,\ell$ be such that $V_{i j \ell}>0$ and $V_{(i-1) j \ell}>0$ (i.e., by \eqref{eq:Vijk}, $0 \le \ell \le m$, $\ell<i\le m$ and $\ell \le j \le m+\ell-i$).
Then 
\begin{align} \label{eq:Vijk-ratio-i}
\frac{V_{i j \ell}}{V_{(i-1) j \ell}} &= \frac{2^{i-\ell}-2^{-\ell}}{2^{i-\ell}-1} \cdot \frac{2^{m+\ell-i+1}-2^j}{2^{m-i+1}-1}.
\end{align}
Moreover, if $\ell >0$,
\begin{align}
\label{eq:Vijk-ratio-i-2}
\frac{V_{i j \ell}}{V_{(i-1) j \ell}}  > 2^\ell  \frac{2^{m-i+1}-2^{j-\ell}}{2^{m-i+1}-1}.
\end{align}
Further, if $j-\ell \le m-i$ and $\ell>0$, 
\begin{align}
\label{eq:Vijk-ratio-i-3}
\frac{V_{i j \ell}}{V_{(i-1) j \ell}}  > 2^{\ell-1}.
\end{align}
}
%

{\lemma \label{lemma:4} Let $i,j,\ell$ be such that $V_{i j \ell}>0$ and $V_{(i-1) j (\ell-1)}>0$ (i.e., by \eqref{eq:Vijk}, $0 < \ell \le i\le m$ and $\ell < j \le m+\ell-i$).
Then 
\begin{align}
\label{eq:Vijk-ratio-i-l}
\frac{V_{i j \ell}}{V_{(i-1) j (\ell-1)}} = \frac{2^i-1 }{2^{m+1}-2^i } \cdot \frac{2^{j+1}-2^\ell }{2^\ell-1}.
\end{align}
Moreover, if $i=\ell$ and $j=m$, $\frac{V_{\ell m \ell}}{V_{(\ell-1) m (\ell-1)}} = 1$.
}

\section{Proof of Theorem~\ref{The:Monotonicity}}
\label{app:ProofTh1}


We prove the monotonicity of $\fvec(\yvec;\varepsilon)$ and $\gvec(\xvec)$ by induction. We first prove that the variable node operation in the CCDF form is non-decreasing in $\yvec$. Let $\avec$ and $\bvec$ be two CCDF vectors. The variable node performs the intersection of two random subspaces, $\W=\U\otimes\V$. We define
\begin{equation}
c_{\circ k}= [\avec_{\circ} \boxdot \bvec_{\circ}]_k = \sum_{i=k}^m a_{\circ i}\sum_{j=k}^{m+k-i} V_{i,j,k} b_{\circ j}.
\end{equation}

We can write the CCDF element $c_k$ as
\begin{align}
\label{eq:ck-1}
c_k & \triangleq\sum_{\ell=k}^m c_{\circ \ell}\\
\label{eq:ck-2}
& =\sum_{\ell=k}^m\sum_{i=\ell}^m a_{\circ i} \sum_{j=\ell}^{m+\ell-i} V_{i,j,\ell}b_{\circ j}
\end{align}

We also remind that (cf.~\eqref{eq:Map2})
\begin{equation}
\label{eq:avecc}
{a_\circ}_i = \begin{cases}
1-a_1, & i=0\\
a_i - a_{i+1}, & 1 \le i <m\\
a_m, & i = m.
\end{cases}
\end{equation}
 
In the following, we prove that ${\derivF {c_k} {a_i}}\ge 0$ by proving that ${\derivF {{c_\circ}_\ell} {a_i}}\ge 0$. Given \eqref{eq:avecc}, there are two cases to treat, $\ell = m$ and $1 \le \ell <m$.

First, consider $\ell = m$. Then,
\begin{align*}
{c_\circ}_m = a_m V_{mmm} {b}_m
\end{align*}
and
\begin{align}
{\derivF {{c_\circ}_m} {a_i}} = \begin{cases}
V_{mmm} {b}_m, & i = m\\
0, & \text{otherwise}.
\end{cases}
\end{align}
Now, assume $1 \le \ell <m$.
We develop
\begin{align}
{c_\circ}_\ell =& \  a_m V_{m \ell \ell} {b_\circ}_\ell + \sum_{i=\ell}^{m-1} \left(a_i - a_{i+1}\right) \sum_{j=\ell}^{\ell+m-i} V_{ij \ell} {b_\circ}_j \\
=& \  a_m V_{m \ell \ell} {b_\circ}_\ell +\sum_{i=\ell}^{m-1} a_i  \sum_{j=\ell}^{\ell+m-i} V_{ij \ell} {b_\circ}_j  \nonumber
\\& \ -\sum_{i=\ell}^{m-1}a_{i+1} \sum_{j=\ell}^{\ell+m-i} V_{ij \ell} {b_\circ}_j \\
=&   \sum_{i=\ell}^{m} a_i  \sum_{j=\ell}^{\ell+m-i} V_{ij \ell} {b_\circ}_j -\sum_{i=\ell}^{m-1}a_{i+1} \sum_{j=\ell}^{\ell+m-i} V_{ij \ell} {b_\circ}_j
\end{align}
\begin{align}
=&   \sum_{i=\ell}^{m} a_i  \sum_{j=\ell}^{\ell+m-i} V_{ij \ell} {b_\circ}_j 
  -\sum_{i=\ell+1}^{m}a_{i} \sum_{j=\ell}^{\ell+m-(i-1)} V_{(i-1)j \ell} {b_\circ}_j \\
= & \sum_{i=\ell}^{m} a_i  \sum_{j=\ell}^{\ell+m-i} V_{ij \ell} {b_\circ}_j  -\sum_{i=\ell+1}^{m}a_{i}\sum_{j=\ell}^{\ell+m-i} V_{(i-1)j \ell} {b_\circ}_j  
\nonumber \\  &
-\sum_{i=\ell+1}^{m}a_{i}V_{(i-1) (\ell+m-i+1) \ell} {b_\circ}_{\ell+m-i+1} .
\end{align}
We continue
\begin{align}
{c_\circ}_\ell =& a_\ell  \sum_{j=\ell}^{m} V_{\ell j \ell} {b_\circ}_j   + \sum_{i=\ell+1}^{m} a_i  \sum_{j=\ell}^{\ell+m-i} \left(V_{i j \ell} - V_{(i-1)j \ell} \right) {b_\circ}_j 
\nonumber \\  &
-\sum_{i=\ell+1}^{m}a_{i}V_{(i-1) (\ell+m-i+1) \ell} {b_\circ}_{\ell+m-i+1} .
\label{eq:co}
\end{align}
Given all above, for all values of $\ell$, $1 \le \ell \le m$, the derivative ${\derivF {{c_\circ}_\ell} {a_i}}$ is
\begin{align}
\label{eq:deriv-cvecc}
{\derivF {{c_\circ}_\ell} {a_i}} = \begin{cases}
0, & i < \ell\\
 \sum_{j=\ell}^{m} V_{\ell j \ell} {b_\circ}_j >0, & i = \ell\\
\eqref{eq:cond3}, & \ell +1 \le i \le m
\end{cases}
\end{align}
where \eqref{eq:cond3} is
\begin{equation}
\label{eq:cond3}
-V_{(i-1) (\ell+m-i+1) \ell} {b_\circ}_{\ell+m-i+1}+ \sum_{j=\ell}^{\ell+m-i} \left(V_{i j \ell} - V_{(i-1)j \ell} \right) {b_\circ}_j.
\end{equation}
Putting together \eqref{eq:ck-1} and \eqref{eq:deriv-cvecc}, we obtain
\begin{equation}
\label{eq:partder}
{\derivF {c_k} {a_i}} = \begin{cases}
0, & i<k;\\
\sum_{j=k}^{m} V_{k j k} {b_\circ}_j >0,  & i=k\\
\sum_{\ell = k}^{i-1} {\derivF {{c_\circ}_\ell} {a_i}} +  {\derivF {{c_\circ}_i} {a_i}} + \sum_{\ell = i+1}^m {\derivF {{c_\circ}_\ell} {a_i}}, & k < i < m\\
\sum_{\ell = k}^{m-1} {\derivF {{c_\circ}_\ell} {a_m}} +  {\derivF {{c_\circ}_m} {a_m}}, & i=m.
\end{cases}\end{equation}
For $k < i <m$,
\begin{align}
 \sum_{\ell = k}^{i-1} {\derivF {{c_\circ}_\ell} {a_i}} +  {\derivF {{c_\circ}_i} {a_i}} + \sum_{\ell = i+1}^m {\derivF {{c_\circ}_\ell} {a_i}}
& = \sum_{\ell = k}^{i-1} {\derivF {{c_\circ}_\ell} {a_i}} +  {\derivF {{c_\circ}_i} {a_i}} + 0.
\end{align}
Therefore, we rewrite \eqref{eq:partder} as
\begin{equation}
{\derivF {c_k} {a_i}} = \begin{cases}
0, & i<k\\
 \sum_{j=k}^{m} V_{k j k} {b_\circ}_j >0,  & i=k\\
\sum_{\ell = k}^{i-1} {\derivF {{c_\circ}_\ell} {a_i}} +  {\derivF {{c_\circ}_i} {a_i}}, & k < i \le m.
\end{cases}\end{equation}
To show that ${\derivF {c_k} {a_i}} $ is nonnegative, one should prove it for the case when $1 \le k<i \le m$.
We have
\begin{multline}
{\derivF {c_k} {a_i}}= \sum_{\ell = k}^{i-1} \left(-  V_{(i-1) (\ell+m-i+1) \ell} {b_\circ}_{\ell+m-i+1}
\right. \\ \left.
+ \sum_{j=\ell}^{\ell+m-i} \left(V_{i j \ell} - V_{(i-1)j \ell} \right) {b_\circ}_j \right) 
 + \sum_{j=i}^{m} V_{i j i} {b_\circ}_j .
\end{multline}
Letting $\sum_{j=i}^{m} V_{i j i} {b_\circ}_j  = \sum_{\ell=i}^{m} V_{i \ell i} {b_\circ}_\ell $, 
\begin{multline}
{\derivF {c_k} {a_i}}  = 
\sum_{\ell=i}^{m} V_{i \ell i} {b_\circ}_\ell -  \sum_{\ell = k}^{i-1} V_{(i-1) (\ell+m-i+1) \ell} {b_\circ}_{\ell+m-i+1}
\\
+ \sum_{\ell = k}^{i-1} \sum_{j=\ell}^{\ell+m-i} \left(V_{i j \ell} - V_{(i-1)j \ell}  \right) {b_\circ}_j.
\end{multline}
By substituting $\ell'=\ell+m-i+1$ in the second summation, 
\begin{multline} \label{eq:deriv-1}
{\derivF {c_k} {a_i}}  = 
\sum_{\ell=i}^{m} V_{i \ell i} {b_\circ}_\ell -  \sum_{\ell' = m+k-i+1}^{m} V_{(i-1) \ell' (\ell'+i-m-1)} {b_\circ}_{\ell'} 
\\ + \sum_{\ell = k}^{i-1} \sum_{j=\ell}^{\ell+m-i} \left(V_{i j \ell} - V_{(i-1)j \ell}  \right) {b_\circ}_j.
\end{multline}

Now, depending on the value of $i$, we have two cases to consider:
$i \le m+k-i+1$ and $i > m+k-i+1$.

\subsection{Case $i \le m+k-i+1$}

Consider $ \sum_{\ell = k}^{i-1} \sum_{j=\ell}^{\ell+m-i} \left(V_{i j \ell} - V_{(i-1)j \ell}  \right) {b_\circ}_j$ and let exchange the sums in $\ell$ and $j$. 
It can be verified that
\begin{itemize}
\item if $k \le j < i-1$, then $\ell$ varies from $k$ to $j$;
\item if $i-1 \le j \le k+m-i$, then $\ell$ varies from $k$ to $i-1$;
\item if $k+m-i < j \le m-1$, then $\ell$ varies from $i+j-m$ to $i-1$.
\end{itemize} 

Therefore, \eqref{eq:deriv-1} can be written as
\begin{multline}
{\derivF {c_k} {a_i}}  = 
\sum_{\ell=i}^{m} V_{i \ell i} {b_\circ}_\ell -  \sum_{\ell = m+k-i+1}^{m} V_{(i-1) \ell (\ell+i-m-1)} {b_\circ}_{\ell} \\
+ \sum_{j=k}^{i-2}  {b_\circ}_j \sum_{\ell = k}^j  \left(V_{i j \ell} - V_{(i-1)j \ell}  \right) 
 + \sum_{j=i-1}^{k+m-i}  {b_\circ}_j \sum_{\ell = k}^{i-1}  \left(V_{i j \ell} - V_{(i-1)j \ell}  \right) \\
+  \sum_{j=k+m-i+1}^{m-1}  {b_\circ}_j \sum_{\ell = i+j-m}^{i-1}  \left(V_{i j \ell} - V_{(i-1)j \ell}  \right).
\end{multline}
By rearranging terms in ${b_\circ}_{m+k-i+1}, \ldots, {b_\circ}_m$, we obtain
\begin{multline}
{\derivF {c_k} {a_i}}  = 
\sum_{\ell=i}^{k+m-i} V_{i \ell i} {b_\circ}_\ell+ \left(V_{i m i}  -  V_{(i-1) m (i-1)} \right) {b_\circ}_{m} \\ + \sum_{j=k}^{i-2}  {b_\circ}_j \sum_{\ell = k}^j  \left(V_{i j \ell} - V_{(i-1)j \ell}  \right)\\
 + \sum_{j=i-1}^{k+m-i}  {b_\circ}_j \sum_{\ell = k}^{i-1}  \left(V_{i j \ell} - V_{(i-1)j \ell}  \right) \\
+  \sum_{j=k+m-i+1}^{m-1}  {b_\circ}_j \left( V_{i \ell i} - V_{(i-1) j (j+i-m-1)}
\right. \\ \left.
+ \sum_{\ell = i+j-m}^{i-1}  \left(V_{i j \ell} - V_{(i-1)j \ell} \right) \right)
\end{multline}
Note that $V_{i m i}  -  V_{(i-1) m (i-1)} = 0$ by Lemma~\ref{lemma:4}.
Thus,
\begin{align}
{\derivF {c_k} {a_i}}  & = 
\sum_{\ell=i}^{k+m-i} V_{i \ell i} {b_\circ}_\ell+ \sum_{j=k}^{i-2}  {b_\circ}_j \sum_{\ell = k}^j  \left(V_{i j \ell} - V_{(i-1)j \ell}  \right) 
\nonumber \\ & ~~~~
\nonumber + \sum_{j=i-1}^{k+m-i}  {b_\circ}_j \sum_{\ell = k}^{i-1}  \left(V_{i j \ell} - V_{(i-1)j \ell}  \right)
\\ & \nonumber
 ~~~~+  \sum_{j=k+m-i+1}^{m-1}  {b_\circ}_j \left( V_{i \ell i}- V_{(i-1) j (j+i-m-1)}
\right. \\ & \left.  
~~~+ \sum_{\ell = i+j-m}^{i-1}  \left(V_{i j \ell} - V_{(i-1)j \ell} \right) \right)
\\&= A_1+A_2+A_3+A_4
\end{align}
where 
\begin{align}
\label{eq:A1} A_1 & = \sum_{\ell=i}^{k+m-i} V_{i \ell i} {b_\circ}_\ell\\
A_2 & = \sum_{j=k}^{i-2}  {b_\circ}_j \sum_{\ell = k}^j  \left(V_{i j \ell} - V_{(i-1)j \ell}  \right)\\
A_3 & =  \sum_{j=i-1}^{k+m-i}  {b_\circ}_j \sum_{\ell = k}^{i-1}  \left(V_{i j \ell} - V_{(i-1)j \ell}  \right)
\end{align}
\begin{align}
A_4 & =   \sum_{j=k+m-i+1}^{m-1}  {b_\circ}_j \left( V_{i \ell i}- V_{(i-1) j (j+i-m-1)}
\nonumber \right. \\ & \left. 
~~~~+ \sum_{\ell = i+j-m}^{i-1}  \left(V_{i j \ell} - V_{(i-1)j \ell} \right) \right).
\end{align}

Clearly, $A_1\ge 0$. Let us now show that $A_2$ and $A_3$ are nonnegative. We have $\ell>0$, and $j-\ell \le i-1-k$ and $i \le m+k-i+1$, so that $j-\ell \le m-i$. Therefore, using \eqref{eq:Vijk-ratio-i-3} in Lemma~\ref{lemma:3},
\begin{align}
A_2               \ge \sum_{j=k}^{i-2}  {b_\circ}_j \sum_{\ell = k}^j V_{(i-1)j \ell} \left(2^{\ell-1}- 1  \right)\ge 0.
\end{align}
Similarly,
\begin{align}
A_3   & \ge \sum_{j=i-1}^{k+m-i}  {b_\circ}_j \sum_{\ell = k}^{i-1}  V_{(i-1)j \ell} \left(2^{\ell-1}  -1  \right)\ge 0.
\end{align}

We show now that $A_4>0$.
For this, consider the term $ V_{i \ell i}  - V_{(i-1) j (j+i-m-1)}+ \sum_{\ell = i+j-m}^{i-1}  \left(V_{i j \ell} - V_{(i-1)j \ell} \right)$. 
Remind that $\ell \ge i+j-m$.
Therefore, by Lemma~\ref{lemma:3},
\begin{align}
V_{i j \ell} - & V_{(i-1)j \ell}  \ge V_{(i-1)j \ell} \left(2^\ell  \frac{2^{m-i+1}-2^{j-\ell}}{2^{m-i+1}-1}- 1  \right)\\
& \ge V_{(i-1)j \ell} \left(2^\ell  \frac{2^{m-i+1}-2^{j-(i+j-m)}}{2^{m-i+1}-1}- 1  \right)\\
& > V_{(i-1)j \ell} (2^{\ell-1}-1).
\end{align}
Then the expression of interest is lower bounded as
\begin{multline}
 V_{i \ell i} - V_{(i-1) j (j+i-m-1)}+ \sum_{\ell = i+j-m}^{i-1}  \left(V_{i j \ell} - V_{(i-1)j \ell} \right) \\>  V_{i \ell i} - V_{(i-1) j (j+i-m-1)}+ \sum_{\ell = i+j-m}^{i-1} V_{(i-1)j \ell} (2^{\ell-1}-1) \\
> V_{i \ell i} - V_{(i-1) j (j+i-m-1)}+ V_{(i-1)j (i+j-m)} (2^{i+j-m-1}-1).
\end{multline}
Consider $$V_{i \ell i}- V_{(i-1) j (j+i-m-1)}+ V_{(i-1)j (i+j-m)} (2^{i+j-m-1}-1).$$ 
We first compute
\begin{align}
&\frac{V_{i \ell i}}{ V_{(i-1) j (j+i-m-1)}} = \frac{{\GaussBin}_{m-i,m-\ell}}{{\GaussBin}_{m,m-\ell}} \frac{{\GaussBin}_{m,m-j} 2^{-(m-j)(m-i+1)}}{{\GaussBin}_{i-1,m-j} }\\
&= \prod_{t_1=0}^{m-\ell-1} \frac{2^{m-i-t_1}-1}{2^{m-t_1}-1} \cdot  \prod_{t_2=0}^{m-j-1} \frac{2^{m-t_2}-1}{2^{m-t_2}-2^{m-i+1}}\\
&\ge \prod_{t_1=0}^{m-\ell-1} \frac{2^{m-i-t_1}-1}{2^{m-t_1}-1} \cdot  \frac{2^{m}-1}{2^{m}-2^{m-i+1}}\\
&\ge 2^{(i-1)(m-\ell-1)} \frac{2^{m}-1}{2^{m}-2^{m-i+1}}\\
&\ge 2^{(m-\ell-1)} \frac{2^{m}-1}{2^{m}-2^{m-i+1}}\\
&\ge 2^{(m-i)} \\
& \ge 2^{m/2} >1.
\end{align}
Hence, we have $V_{i \ell i} > V_{(i-1) j (j+i-m-1)}$.
This implies  \\$A_4>0$. 

Finally, putting together $A_1$, $A_2$, $A_3$, and $A_4$, it follows that ${\derivF {c_k} {a_i}} > 0$ for $i \le m+k-i+1$.

\subsection{Case $i > m+k-i+1$}

Consider $ \sum_{\ell = k}^{i-1} \sum_{j=\ell}^{\ell+m-i} \left(V_{i j \ell} - V_{(i-1)j \ell}  \right) {b_\circ}_j$ and let exchange the sums in $\ell$ and $j$. It can be verified that,
\begin{itemize}
\item if $k \le j \le k+m-i$, then $\ell$ varies from $k$ to $j$;
\item if $k+m-i < j \le i-1$, then $\ell$ varies from $i+j-m$ to $j$;
\item if $i-1 < j \le m-1$, then $\ell$ varies from $i+j-m$ to $i-1$.
\end{itemize} 
Thus, \eqref{eq:deriv-1} is written as
\begin{multline}
{\derivF {c_k} {a_i}}  = 
\sum_{\ell=i}^{m} V_{i \ell i} {b_\circ}_\ell -  \sum_{\ell = m+k-i+1}^{m} V_{(i-1) \ell (\ell+i-m-1)} {b_\circ}_{\ell} \\
+ \sum_{j=k}^{k+m-i}  {b_\circ}_j \sum_{\ell = k}^j  \left(V_{i j \ell} - V_{(i-1)j \ell}  \right) \\
 + \sum_{j=k+m-i+1}^{i-1}  {b_\circ}_j \sum_{\ell = i+j-m}^{j}  \left(V_{i j \ell} - V_{(i-1)j \ell}  \right)
\\+  \sum_{j=i}^{m-1}  {b_\circ}_j \sum_{\ell = i+j-m}^{i-1}  \left(V_{i j \ell} - V_{(i-1)j \ell}  \right).
\end{multline}
By rearranging terms in ${b_\circ}_{m+k-i+1}, \ldots, {b_\circ}_m$, we obtain
\begin{align}
{\derivF {c_k} {a_i}} & = 
\left(V_{i m i}  -  V_{(i-1) m (i-1)} \right) {b_\circ}_{m}
\nonumber \\ &
~~~+ \sum_{j=k}^{k+m-i}  {b_\circ}_j \sum_{\ell = k}^j  \left(V_{i j \ell} - V_{(i-1)j \ell}  \right) \nonumber\\
& ~~~ + \sum_{j=k+m-i+1}^{i-1}  {b_\circ}_j \left(V_{i \ell i} - V_{(i-1) \ell (\ell+i-m-1)}
\nonumber  \right. \\   & \left.
~~~+\sum_{\ell = i+j-m}^{j}  \left(V_{i j \ell} - V_{(i-1)j \ell}  \right) \right)
\nonumber \\&~~~ +  \sum_{j=i}^{m-1}  {b_\circ}_j \left(V_{i \ell i} - V_{(i-1) \ell (\ell+i-m-1)}
\nonumber  \right. \\   & \left.
~~~+\sum_{\ell = i+j-m}^{i-1}  \left(V_{i j \ell} - V_{(i-1)j \ell}  \right) \right).
\end{align}
As before, $V_{i m i}  -  V_{(i-1) m (i-1)} = 0$ by Lemma~\ref{lemma:4}.
Hence
\begin{align}
{\derivF {c_k} {a_i}} =  B_1 + B_2 + B_3
\end{align}
where
\begin{align}
B_1 & =  \sum_{j=k}^{k+m-i}  {b_\circ}_j \sum_{\ell = k}^j  \left(V_{i j \ell} - V_{(i-1)j \ell}  \right)
\end{align}
\begin{align}
B_2&=\sum_{j=k+m-i+1}^{i-1}  {b_\circ}_j \left(V_{i \ell i} - V_{(i-1) \ell (\ell+i-m-1)} \right. \nonumber
\\ & \left. ~~~~+\sum_{\ell = i+j-m}^{j}  \left(V_{i j \ell} - V_{(i-1)j \ell}  \right) \right)\\
B_3 &=  \sum_{j=i}^{m-1}  {b_\circ}_j \left(V_{i \ell i} - V_{(i-1) \ell (\ell+i-m-1)} \right. \nonumber
\\ & \left. ~~~~+\sum_{\ell = i+j-m}^{i-1}  \left(V_{i j \ell} - V_{(i-1)j \ell}  \right) \right).
\end{align}
We need to show that $B_1$, $B_2$ and $B_3$ are nonnegative. 
For $B_1$, $j-\ell \le m-i$, hence, by Lemma~\ref{lemma:3}, $B_1\ge 0$.
The fact that $B_2 > 0$ and $B_3 > 0$ is proven exactly in the same way as for $A_4$.

Finally, given that $B_1$, $B_2$ and $B_3$ are nonnegative, ${\derivF {c_k} {a_i}} > 0$ for $i > m+k-i+1$.

For the general variable node operation $\boxdot^{\dl-1}$, we do the following. Consider the implicitly defined vector
function $\cvec=\hvec(\avec,\bvec)$ given by \eqref{eq:ck-2}. Using this, we define the recursion $\hvec^{t}(\avec,\bvec) =\hvec(\avec,\hvec^{t-1}(\avec,\bvec))$ starting from $\hvec^1(\avec,\bvec) = \hvec(\avec,\bvec)$. 
From the fact that $c_k$ is a non-decreasing function of $a_i$ and $b_i$ it follows that the Jacobian derivatives $\derivF{\hvec(\avec,\bvec)}{\avec}$ and $\derivF{\hvec(\avec;\bvec)}{\bvec}$ are nonnegative matrices. 
Using the recursion defined above,
one can show that the Jacobian derivative of $\hvec^t(\avec)$ is a nonnegative
matrix because it is the sum of products of nonnegative
matrices. Therefore, $\hvec^{\dl-1}(\avec,\bvec)$ is increasing in $\avec$. Hence, $\fvec(\yvec;\varepsilon)$ is increasing in $\yvec$, i.e., if $\yvec_1\preceq \yvec_2$ then $\fvec(\yvec_1;\varepsilon)\preceq \fvec(\yvec_2;\varepsilon)$ 

The proof for the monotonicity of $\gvec(\xvec)$ follows the same lines.

\section{Derivative of $U(\xvec;\varepsilon)$}
\label{app:DerivativeU}

We compute the derivative of $U(\xvec;\varepsilon)$ with respect to $\xvec$,
\begin{equation}
\label{eq:UprimeApp}
U'(\xvec;\varepsilon)=\deriv{\xvec}  \left(\gvec(\xvec)\Dvec \xvec\T -G(\xvec)-F(\gvec(\xvec);\varepsilon)\right)
\end{equation}

We consider separately the derivatives of the three terms,
\begin{enumerate}
\item 
\begin{align}
\label{eq:d1}
\deriv{\xvec}\left(\gvec(\xvec)\right)\left(\Dvec\xvec\T\right)&=\gvec(\xvec)\derivF{\Dvec\xvec\T}{\xvec}+\left(\Dvec\xvec\T\right)\T\derivF{\gvec(\xvec)}{\xvec}\nonumber\\
&=\gvec(\xvec)\Dvec+\xvec\Dvec\T\gvec'(\xvec)
\end{align}
\item 
\begin{align}
\label{eq:d2}
\derivF{G}{\xvec}=\gvec(\xvec)\Dvec
\end{align}
\item\begin{align}
\label{eq:d3}
\derivF{F}{\xvec}=\fvec(\gvec(\xvec);\varepsilon)\Dvec\gvec'(\xvec)
\end{align}
\end{enumerate}
Assuming $\Dvec$ symmetric, i.e., $\Dvec=\Dvec\T$, substituting (\ref{eq:d1})--(\ref{eq:d3}) in (\ref{eq:UprimeApp}) we obtain (\ref{eq:DerU}).

\section{Proof of Theorem~\ref{Th:Uprime}}
\label{app:ProofTh2}

We compute the partial derivative of the three terms in (\ref{eq:PotFunctionVect}).

1) Derivative of $\tr(\Gvec(\Xvec) \Dvec \Xvec\T )$.
First note that
\begin{equation} \label{eq:1}
\tr(\Gvec(\Xvec) \Dvec \Xvec\T ) = \sum_{j=1}^L \left[ \Gvec(\Xvec) \Dvec \right]_j \xvec_j\T = \sum_{j=1}^L  \gvec(\xvec_j)   \Dvec \xvec_j\T.\end{equation}
We will use the following lemma.
{\lemma \label{lemma:deriv} Let $\gvec(\xvec_i)$ and $\svec(\xvec_i)$ be two $1\times m$ vectors. Define
\begin{align}
{\derivF {\gvec(\xvec_i)} {\xvec_i}} &= \left( {\derivF {[\gvec(\xvec_i)]_k}{x_{i\ell}}} \right)_{k=1,\ldots,m,~\ell=1,\ldots,m}\nonumber\\
\quad {\derivF {\svec(\xvec_i)} {\xvec_i}} &= \left( {\derivF {[\svec(\xvec_i)]_k}{x_{i\ell}}} \right)_{k=1,\ldots,m,~\ell=1,\ldots,m}.\nonumber \end{align}
Then,
\begin{equation} \nonumber
{\deriv {\xvec_i} \gvec(\xvec_i) \svec(\xvec_i)\T} = \svec(\xvec_i) {\derivF {\gvec (\xvec_i)} {\xvec_i}} + \gvec(\xvec_i){\derivF {\svec(\xvec_i)\T} {\xvec_i}}.
\end{equation}
}
The proof of the lemma is omitted for brevity.

Applying Lemma~\ref{lemma:deriv} to (\ref{eq:1}) with $\gvec(\xvec_j)$ and $\svec(\xvec_j)=(\Dvec \xvec_j\T)\T=\xvec_j\Dvec\T$, and taking into account that
$${\derivF {\Dvec \xvec_j\T} {\xvec_j}} = \left({\derivF {[\Dvec\xvec_j\T]_k} {x_{j\ell}}} \right)_{k=1,\ldots,m,~\ell=1,\ldots,m} = 
\Dvec$$
we obtain
\begin{equation}\nonumber
{\deriv {\xvec_j} } \tr(\Gvec(\Xvec) \Dvec \Xvec\T ) =  \xvec_j \Dvec \Gvec_{\mathrm d}(\xvec_j) + \gvec(\xvec_j) \Dvec.
\end{equation}
where we have used the fact that $\Dvec\T=\Dvec$ if $\Dvec$ is symmetric.

2) Derivative of $G(\Xvec)$. It is easy to see that $\derivF {G(\Xvec)} {\xvec_j} = \gvec(\xvec_j) \Dvec$.  

3) Derivative of $F(\Avec \Gvec(\Xvec);\varepsilon)$. Let $\Yvec=\Avec \Gvec (\Xvec)$. Then
\begin{align}
{\deriv \xvec_i} F(\Yvec; \varepsilon)
&= \sum_{j=1}^L  {\deriv \xvec_i} F( \yvec_j ; \varepsilon) \nonumber\\
&= \sum_{j=1}^L \left[{\deriv {{\xvec}_{i1}}} F( \yvec_j ; \varepsilon),\ldots,{\deriv {{\xvec}_{im}}} F( \yvec_j ; \varepsilon)\right]\nonumber
\end{align}

Note that
\begin{equation}
\label{eq:derivj}
{\derivF {F(\yvec_j; \varepsilon)} {x_{i\ell}}} = \sum_{n=1}^m {\derivF {F(\yvec_j; \varepsilon)} {y_{jn}} } {\derivF {y_{jn}} {x_{i\ell}}}.
\end{equation}
Define the matrix $\Bvec={\derivF {F(\Yvec; \varepsilon)} {\yvec}}$ such that its $(j,m)$-th entry $[\Bvec]_{j,n}=B_{jn}={\derivF {F(\yvec_j; \varepsilon)} {y_{jn}} }$. By definition $\Bvec= \fvec(\Yvec; \varepsilon)\Dvec$ and $B_{jn}= \left[\fvec(\yvec_j; \varepsilon) \Dvec\right]_{j,n}$. Now let us find ${\derivF {y_{jn}} {x_{i\ell}}}$.
As
$$y_{jn} = [\Avec\Gvec(\Xvec)]_{j,n} =\sum_{t=1}^L A_{jt} [\Gvec (\Xvec
)]_{t,n}$$
we can write
\begin{align}
{\derivF {y_{jn}} {x_{il}}} &=
\sum_{t=1}^L A_{jt} {\derivF {[\Gvec (\Xvec)]_{tn}}{x_{i\ell}}}\nonumber\\
&\stackrel{(a)}{=} A_{ji} {\derivF {[\Gvec (\Xvec)]_{in}}{x_{i\ell}}}\stackrel{(b)}{=} A_{ji} \left[ \Gvec_{\mathrm d}(\xvec_i)\right]_{n,\ell}\nonumber
\end{align}
where $A_{jt}=[\Avec]_{j,t}$ and $(a)$ follows from the fact that the only non-zero term in the sum over $t$ is for $t=i$, and $(b)$ follows from the definition of $\Gvec_{\mathrm d}(\xvec_i)$, $\Gvec_{\mathrm d}(\xvec_i)= \left( {\derivF {[\gvec(\xvec_i)]_k} {x_{i\ell}}} \right)_{k=1,\ldots,m, \ \ell=1,\ldots,m}$. Therefore, (\ref{eq:derivj}) becomes
\begin{align}
{\derivF {F_j(\yvec_j; \varepsilon)} {x_{i\ell}}} &= A_{ji}  \sum_{n=1}^m B_{jn}  \left[ \Gvec_{\mathrm d}(\xvec_i)\right]_{n,l}\nonumber
\end{align}
and
\begin{align}
&{\deriv \xvec_i} F(\Yvec; \varepsilon)
= \sum_{j=1}^L  \left[{\deriv {{\xvec}_{i1}}} F( \yvec_j ; \varepsilon),\ldots,{\deriv {{\xvec}_{im}}} F( \yvec_j ; \varepsilon)\right]\nonumber\\
&=\sum_{j=1}^L A_{ji} \left[\sum_{n=1}^m B_{jn}  \left[ \Gvec_{\mathrm d}(\xvec_i)\right]_{n,1},\ldots,\sum_{n=1}^m B_{jn}  \left[ \Gvec_{\mathrm d}(\xvec_i)\right]_{n,m}\right]\nonumber\\
&=\sum_{j=1}^L A_{ji}[\Bvec]_j\Gvec_{\mathrm d}(\xvec_i)
=\left[\Avec\T\right]_i\Bvec\Gvec_{\mathrm d}(\xvec_i).\nonumber
\end{align}

Finally, we obtain
\begin{align}
{\deriv \xvec_i} F(\Avec\Gvec(\Xvec); \varepsilon)
&=\left[\Avec\T\right]_i\Fvec(\Avec\Gvec(\Xvec); \varepsilon)\Dvec\Gvec_{\mathrm d}(\xvec_i).\nonumber
\end{align}

\bibliographystyle{IEEEtran}
\bibliography{threshold-saturation}

\end{document}